\documentclass[11pt,a4paper]{article}
\usepackage{jheppub}
\usepackage{slashed}
\usepackage{braket}
\usepackage{mma}
\usepackage{amsmath}
\usepackage{amssymb}
\usepackage{xspace}
\usepackage{subcaption}

\newcommand{\eq}[1]{eq.~\eqref{eq:#1}}
\newcommand{\eqs}[2]{eqs.~\eqref{eq:#1} and \eqref{eq:#2}}
\renewcommand{\sec}[1]{section~\ref{sec:#1}}
\newcommand{\secs}[2]{sections~\ref{sec:#1} and \ref{sec:#2}}
\newcommand{\app}[1]{app.~\ref{app:#1}}
\newcommand{\fig}[1]{figure~\ref{fig:#1}}

\newcommand{\nn}{\nonumber}

\newcommand{\df}{\mathrm{d}}
\newcommand{\img}{\mathrm{i}}

\newcommand{\Li}{\textrm{Li}}

\newcommand{\al}{\alpha}
\newcommand{\bt}{\beta}

\newcommand{\de}{\delta}
\newcommand{\eps}{\epsilon}

\newcommand{\si}{\sigma}

\newcommand{\cJ}{{\mathcal J}}
\newcommand{\cO}{{\mathcal O}}

\newcommand{\Mathematica}{\textsc{Mathematica}\xspace}
\newcommand{\Vegas}{\textsc{Vegas}\xspace}
\newcommand{\gojet}{\textsc{GOJet}\xspace}

\allowdisplaybreaks[2]

\preprint{\vbox{
\hbox{Nikhef 20-009}}}

\title{One-loop Jet Functions by Geometric Subtraction}

\author[a]{Avanish Basdew-Sharma,}
\author[a,b]{Franz Herzog,}
\author[a,c]{Solange Schrijnder van Velzen,}
\author[a,c]{Wouter~J.~Waalewijn}
\affiliation[a]{Nikhef, Theory Group, Science Park 105, 1098 XG, Amsterdam, The Netherlands}
\affiliation[b]{Higgs Centre for Theoretical Physics, School of Physics and Astronomy, The University of Edinburgh, Edinburgh EH9 3FD, Scotland, UK}
\affiliation[c]{Institute for Theoretical Physics Amsterdam and Delta Institute for Theoretical Physics, University of Amsterdam, Science Park 904, 1098 XH Amsterdam, The Netherlands}
\emailAdd{avanishb@nikhef.nl}
\emailAdd{fherzog@ed.ac.uk}
\emailAdd{s.v.schrijndervanvelzen@uva.nl}
\emailAdd{w.j.waalewijn@uva.nl}

\abstract{
In factorization formulae for cross sections of scattering processes, final-state jets are described by jet functions, which are a crucial ingredient in the resummation of large logarithms. We present an approach to calculate generic one-loop jet functions, by using the geometric subtraction scheme. This method leads to local counterterms generated from a slicing procedure; and whose analytic integration is particularly simple. The poles are obtained analytically, up to an integration over the azimuthal angle for the observable-dependent soft counterterm. The poles depend only on the soft limit of the observable, characterized by a power law, and the finite term is written as a numerical integral. We illustrate our method by reproducing the known expressions for the jet function for angularities, the jet shape, and jets defined through a cone or $k_{T}$ algorithm. As a new result, we obtain the one-loop jet function for an angularity measurement in $e^+e^-$ collisions, that accounts for the formally power-suppressed but potentially large effect of recoil. An implementation of our approach is made available as the \gojet \Mathematica package accompanying this paper.
}

\begin{document}
\maketitle

\section{Introduction}
\label{sec:intro}

Experimental studies at the Large Hadron Collider (LHC) impose restrictions on QCD radiation in the final state, to stress test the Standard Model and search for New Physics. If these restrictions are tight, they lead to large logarithms in the corresponding cross section. For example, for Higgs plus one jet production with a veto on additional jets with transverse momentum above $p_T^{\rm veto}$, the cross section takes the following form 
\begin{align} \label{eq:si_ex}
\si(p_T^{\rm veto}) = \si_0 \biggl[1 + \sum_{\substack{n\geq 1 \\ 2n \geq m\geq 0}} c_{n,m}\, \al_s^{n} \ln^m \Bigl(\frac{m_H}{p_T^{\rm veto}}\Bigr) + \mathcal{O}\Bigl(\frac{p_T^{\rm veto}}{m_H}\Bigr)\biggr]
\,,\end{align}
where $\si_0$ is the leading-order cross section, and the coefficients $c_{n,m}$ are independent of $p_T^{\rm veto}$. For a tight veto $p_T^{\rm veto} \ll m_H\sim p_T^{\text{jet}}$, the expansion in $\al_s$ deteriorates due to the large logarithms and resummation is crucial to improve convergence and reduce the theory uncertainty. Resummation captures the dominant effect of higher-order corrections, effectively treating $\ln (m_H/p_T^{\rm veto}) \sim 1/\al_s$. 

Large logarithms arise because the cross section involves multiple scales that are widely separated. Resummation of these logarithms can be achieved by factorizing the cross section into components that each involve a single scale, using diagrammatic methods in QCD, see e.g.~\cite{
Dokshitzer:1978hw,Parisi:1979se,Curci:1979bg,Collins:1981uk,Kodaira:1981nh,Bodwin:1984hc,Collins:1984kg,Collins:1989gx}, or Soft-Collinear Effective Theory (SCET)~\cite{Bauer:2000ew, Bauer:2000yr, Bauer:2001ct, Bauer:2001yt,Beneke:2002ph}. For exclusive Higgs plus one jet production, discussed in \eq{si_ex}, this takes on the following (schematic) form $\si \sim HSBBJ$~\cite{Liu:2012sz,Liu:2013hba}. The hard function $H$ describes hard scattering, the soft function $S$ encodes the effect of soft radiation, and the beam functions $B$ and jet function $J$ account for initial- and final-state collinear radiation. The structure of this factorization not only depends on the process, but also on the observable and can involve convolutions between ingredients (though it is simply a product in the above example). Because each ingredient in the factorization involves a single scale, the large logarithms can be resummed by evaluating each ingredient at its natural scale and using the renormalization group to evolve them to a common scale. Alternatively, an automated approach to resummation was pursued in refs.~\cite{Banfi:2004yd,Banfi:2014sua}.

In this paper we focus on calculating one-loop jet functions, which enter in resummed cross sections starting at next-to-leading logarithmic (NLL$'$) accuracy. Resummation at NLL$'$ includes the two-loop cusp anomalous dimension and one-loop (non-cusp) anomalous dimensions.
Jet functions have been calculated for a wide range of observables, including the invariant mass~\cite{Bauer:2003pi,Becher:2006qw,Becher:2009th,Becher:2010pd,Bruser:2018rad,Banerjee:2018ozf}, the family of $e^+e^-$ event shapes called angularities with respect to the thrust axis~\cite{Hornig:2009vb,Becher:2012qc,Bell:2018gce} or Winner-Take-All axis~\cite{Larkoski:2014uqa,Procura:2018zpn}, Sterman-Weinberg jets~\cite{Jouttenus:2009ns,Chay:2015ila}, the cone and the $k_T$ family of jet algorithms for exclusive~\cite{Ellis:2010rwa,Chay:2015ila} and inclusive~\cite{Kang:2016mcy,Dai:2016hzf} jet production. Jet functions have also been considered for a range of jet substructure observables, such as the jet shape~\cite{Li:2011hy,Chien:2014nsa,Cal:2019hjc}. In our calculations we treat quarks as massless and restrict to infrared-safe observables. An example of a massive quark (initiated) jet function is given in refs.~\cite{Fleming:2007xt,Hoang:2019fze}, and an example of an infrared-unsafe jet observable is the electric charge of the jet~\cite{Krohn:2012fg,Waalewijn:2012sv}.

We briefly comment on the other ingredients in the factorization: A general approach to calculating soft functions has been developed in refs.~\cite{Kasemets:2015uus,Bell:2015lsf,Bell:2018oqa,Bell:2020yzz}. In particular,  the \textsc{SoftSERVE} package~\cite{Bell:2020yzz} provides two-loop soft functions for processes with two collinear directions (i.e.~two jets in $e^+e^-$ or 0 jets in $pp$ collisions), and an extension to $N$ jets is in progress~\cite{Bell:2018mkk}. Hard functions can be obtained from the IR finite part of helicity amplitudes, as long as the color of the initial (final) particles is not averaged (summed) over, see e.g.~ref.~\cite{Moult:2015aoa}. 

The difficulty in calculating jet functions lies in the phase-space integration, which depends on the observable. When feasible, an analytic approach is superior. However, there are observables for which even the one-loop jet function is highly nontrivial, such as jet broadening~\cite{Becher:2012qc} and the jet shape~\cite{Cal:2019hjc}, for which fully analytic results are difficult to obtain or have not been obtained yet. The numerical approach we develop here offers a promising alternative, addressing the collinear and soft divergences in a general way, thereby automating the calculation of one-loop jet functions for a broad range of observables. At minimum, our work provides a valuable cross check for analytic calculations. 

The poles in the dimensional regulator are obtained analytically, possibly up to an integral over the azimuthal angle, and depend on the collinear and soft behavior of the observable. This soft behavior is described by a power law, and therefore simply characterized by the exponent and coefficient. Extracting these parameters may require solving non-trivial algebraic equations, and we develop a procedure to simplify this step. The full details/complications of the measurement only enter in the finite term, which can be integrated numerically. We have implemented our approach in a \Mathematica package, Geometric One-loop Jet functions (\gojet), which accompanies this paper. \gojet can handle a large class of infrared-safe observables, including all the observables listed above. 

Using \gojet we provide explicit examples of the method for the angularities with respect to the Winner-Take-All axis, the cone and $k_T$-clustering jet algorithms and the jet shape. Furthermore we calculate for the first time the one-loop jet function for angularities with respect to the thrust axis including recoil. We cross check our result against existing results in the literature for the specific case of jet broadening  \cite{Becher:2012qc} and for the case of no recoil \cite{Hornig:2009vb,Budhraja:2019mcz}.

The remainder of the paper is structured as follows: In section \ref{sec:method} we discuss how we use geometric subtraction  to calculate jet functions, including a simple example. The \gojet package, which provides a \Mathematica implementation, is discussed in \sec{prog}. In \sec{ex}, we use our package to calculate several one-loop jet functions, and we conclude in \sec{conc}.

\section{General Method}
\label{sec:method}

In \sec{subscheme} we will discuss geometric subtraction and how we apply it to calculate one-loop jet functions. Technical aspects related to the treatment of Heaviside theta functions in our calculation and infrared safety are discussed in \secs{dfs}{IRS}, respectively. We illustrate our method by calculating the jet function for the $e^+e^-$ angularity event shapes in \sec{exang}, with further examples in \sec{ex}.

\subsection{Subtraction scheme}
\label{sec:subscheme}

The jet function depends on the flavor $i = q, g$ of the initiating parton and the jet observable, and has a perturbative expansion in $\al_s$
\begin{align}
  \cJ_{i, \rm obs} = \sum_n \Big(\frac{\al_s}{2\pi}\Big)^n \cJ_{i,\rm obs}^{(n)}
\,.\end{align}
At tree level the jet consists of a single quark or gluon, and in general $\cJ_i^{(0)}=1$ in the appropriate units.\footnote{An exception is the jet shape, discussed in \sec{pedro}, which contains a theta function that sets it to zero if the recoil from soft radiation is too large.}
The one-loop contribution is given by the collinear limit of two final-state partons 
\begin{align}
\label{eq:method1}
\mathcal{J}_{i, \rm obs}^{(1)} &= \int_0^{\pi}\! \df \phi \int_0^\infty\! \df s \int_0^1\! \df z \,Q_{i}(s,z,\phi)
  \, M_{\rm obs}(s,z,\phi)\,, 
  \nn \\
 Q_i(s,z,\phi) &=
\frac{(\mu^2e^{\gamma_E})^{\epsilon}  }{ \sqrt{\pi}\,\Gamma(\tfrac{1}{2}-\epsilon)}\bigg(\frac{\nu}{\omega}\bigg)^{\eta}\frac{P_i(z)\,(\sin\phi)^{-2\eps}}{z^{\epsilon+\eta} (1-z)^{\epsilon+\eta} s^{1+\epsilon}},
\nn \\
 P_q(z) &= C_F \bigg[\frac{1+z^2}{1-z} - \eps (1-z) \bigg],
 \nn \\
  P_g(z) &= n_{f}T_{R}\bigg[1-\frac{2z(1-z)}{1-\eps}\bigg] +  C_{A}\bigg[\frac{z}{1-z} + \frac{1-z}{z} + z(1-z)\bigg]
\,.\end{align}
Here $s$ denotes the invariant mass of the two partons, and $z$ and $1-z$ the momentum fractions of the partons. The squared matrix element is contained in $Q_{i}(s,z,\phi)$, with $P_{i}(z)$ the (sum of) splitting function(s). The calculation is performed in $d = 4 - 2\epsilon$ dimensions and the $\overline{\rm MS}$-renormalization scheme with renormalization scale $\mu$ is employed. For certain observables an additional rapidity regulator $\eta$ and corresponding rapidity scale $\nu$ are required~\cite{Becher:2010tm,Chiu:2011qc,Collins:2011zzd,GarciaEchevarria:2011rb,Becher:2011dz,Chiu:2012ir}, which is included in \eq{method1} for generality. This arises when the collinear and soft functions have the same invariant mass scale $\mu$, with transverse momentum measurements being the typical example. For the extension of \eq{method1} to a two-loop example, see ref.~\cite{Ritzmann:2014mka}.

\begin{figure}[t]
\centering
\includegraphics[width=0.6\textwidth]{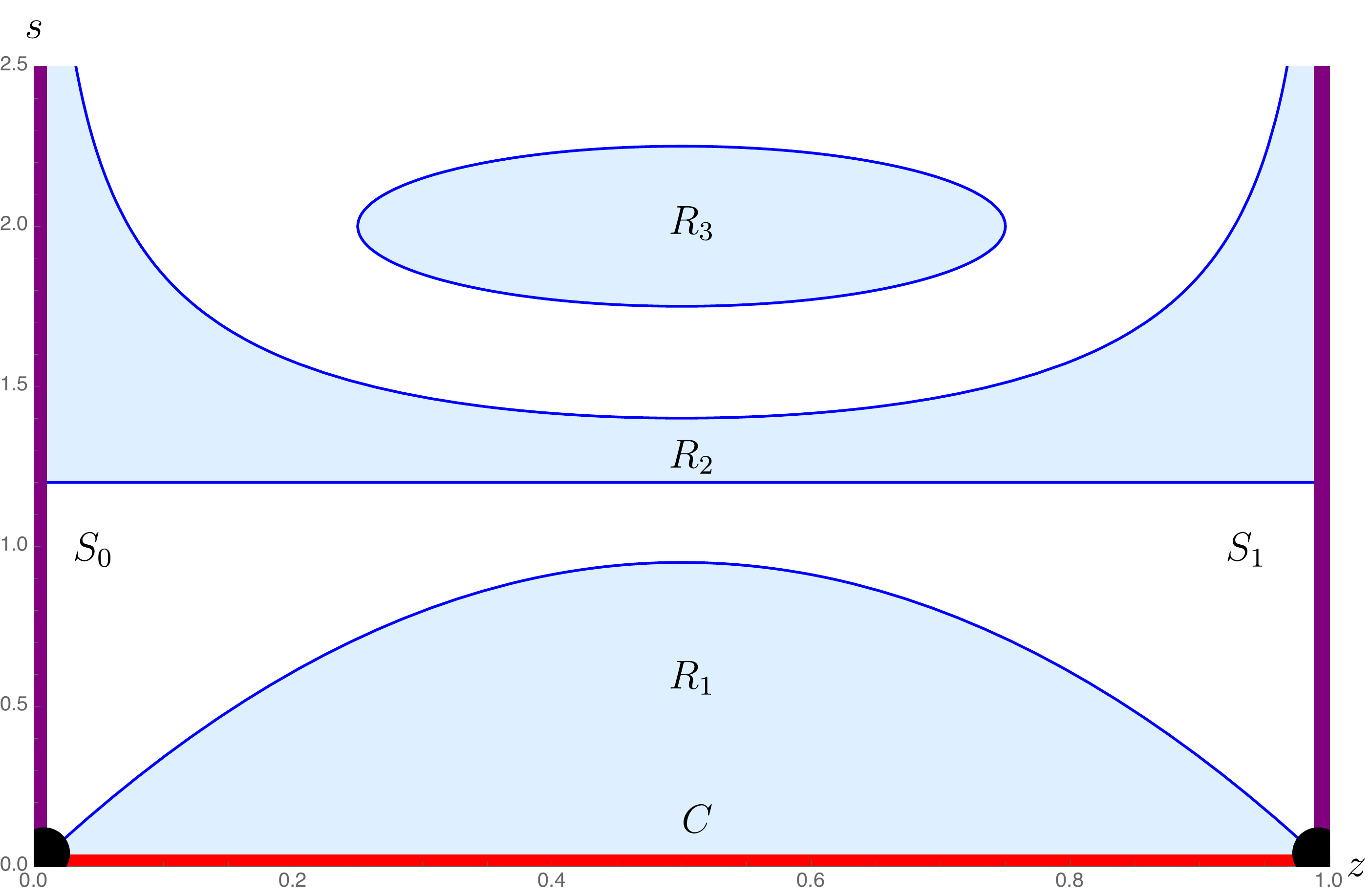}
\caption{For a general observable the phase space can be constrained to several regions (blue). The collinear singularity $C$ (red line), soft singularities $S_{0}$ and $S_{1}$ (purple lines), and soft-collinear singularities (black dots) are indicated.
}
\label{fig:phsp} 
\end{figure}

The measurement in a jet function can often be written as $\de[\cO - f(s,z,\phi)]$. To avoid distributions, we require the user to rewrite the measurement as a Heaviside theta function by integrating, i.e.~$\Theta[\cO - f(s,z,\phi)]$, where we are now cumulative in $\cO$.\footnote{Alternatively, one can consider a conjugate space, as was employed in automated calculations of soft functions~\cite{Bell:2015lsf,Bell:2018oqa}.} We therefore assume that the measurement $M_{\rm obs}(s,z,\phi)$ is a Heaviside theta function, which cuts out a certain region of the collinear phase space, as illustrated in \fig{phsp} (suppressing $\phi$ dependence).
An advantage of cumulative distributions is that they involve logarithms rather than plus distributions: 
\begin{align} \label{eq:cum_rel}
   \int_0^{t^c}\! \df t\, \Big[\frac{\theta(t)\ln^n t}{t}\Big]_+ = \frac{1}{n+1}\, \ln^{n+1} t^c
\,.\end{align}
In \sec{dfs}, a technical point related to rewriting measurement delta functions in terms of theta functions will be discussed.
There are also measurements that are naturally theta functions. For example, the $k_T$-family of jet algorithms requires both particles to be clustered into a jet with radius parameter $R$, $M_{k_T}(s,z,\phi) = \Theta(s\leq z(1-z)p_T^2 R^{2})$, where $p_T$ is the transverse momentum of the jet.
In principle these phase-space constraints $M_{\rm obs}$ can depend on the azimuthal angle $\phi$ as well, but since there is no singularity associated with the $\phi$ integration, we will only include $\phi$ when needed. 

The jet function in \eq{method1} has divergences as  $s \to 0$ (collinear divergence), and $z\to 0$ and $z \to 1$ (soft divergences), which occur at the phase-space boundaries in \fig{phsp}. Infrared-safe observables must always either include or exclude the entire collinear divergence (the red line in \fig{phsp}), as will be discussed more in \sec{IRS}. From the point of view of collinear subtraction, one can consider the jet function (as long as it contains the collinear divergence) as a collinear counterterm. Different observables can then be viewed as different schemes, differing in the extent that soft and soft-collinear divergences are included in the observable. For instance, region 1 of the general observable illustrated in \fig{phsp} only contains the collinear and part of the soft-collinear singularities. By contrast, region 2 only contains part of the soft and none of the collinear divergence.  Region 3 does not contain any soft or collinear divergent parts of phase space and does therefore not have to be regulated. 
Another possibility would be to consider an observable which corresponds to the complement of region 1, which naively causes problems because it develops a logarithmic singularity for $s \to \infty$. However, its one-loop jet function is given by minus the jet function for region 1, because the integral over the full collinear phase space results in a scaleless integral.

To define a general subtraction scheme for calculating jet functions for infrared-safe observables, we follow the approach of geometric subtraction \cite{Herzog:2018ily}.  We would like to define a finite part of the jet function as follows:
\begin{align}
\label{eq:finitepart}
 \mathrm{Finite}(\mathcal{J}_{i, \rm obs}^{(1)}) &= \Big[\int_0^\pi\! \df \phi \int_{B\mu^2}^\infty\! \df s \int_A^{1-A}\! \df z \,Q_{i \to j}(s,z,\phi)\, M_{\rm obs}(s,z,\phi)\Big]_{A,B\to 0} \,,
\end{align}
where we introduced the dimensionless slicing parameters $A$ and $B$, that remove the soft and collinear divergence, and which we subsequently want to take to zero. The central idea of geometric subtraction rests on the identity:
\begin{align}
\label{eq:geosub}
\Big[\int_a^1\! \df x\,\frac{f(x)}{x} \Big]_{a\to0}&=\Big[\int_0^1\! \df x\,\frac{f(x)-f(x)\Theta(x<a)}{x} \Big]_{a\to0}\nn\\
&=\int_0^1\! \df x\,\frac{f(x)-f(0)\Theta(x<a)}{x}\, ,
\end{align}
where we exploited that $a$ is small on the second line to replace $f(x)$ by $f(0)$ in the second term. However, the expression on the second line is now regulated for  any $0<a\le1$, leading to a duality between slicing and subtraction schemes. 
To obtain the full jet function from the above finite part, counterterms need to be added to reinstate the part of the integral that is removed by the cuts. The counterterms generated in this way are added back in integrated form, regulated dimensionally and if needed also with a rapidity regulator, and may give a finite contribution to the jet function.
While a subtlety arises in general when different limits do not commute, here we do not face this problem as the collinear and soft singularities are factorized.
For the small $A$ limit in \eq{finitepart} we can then straightforwardly apply \eq{geosub}. However for the parameter $B$ nothing is gained from this procedure, because the jet function is already in the limit of small $s$ and the counterterm generated is the original integral itself. 

To obtain a simpler counterterm in the $s<B$ region, we can however use a simpler observable, which we choose to be the jet mass, as a collinear counterterm. (This was also used in the geometric subtraction scheme~\cite{Herzog:2018ily}.) Since the region of the $s$-$z$ plane corresponding to the jet mass is box-shaped, we will refer to this collinear counterterm as the \emph{box}. 
A subtlety now appears due to the difference of soft and soft-collinear divergences included in the box counterterm and the given observable $M_{\text{obs}}$, which as discussed above may not be the same. To deal with this problem we introduce separate soft counterterms for both the box counterterm and the $M_{\text{obs}}$ term in the region $s<B\mu^2$, as discussed in detail below. 

These considerations lead us to the following final decomposition of the jet-function into finite and divergent parts:
\begin{align} \label{eq:decomp}
  \mathcal{J}_{i, \rm obs}^{(1)} &= G_{i, {\rm obs},1} + G_{i, {\rm obs},2} + G_{i, {\rm obs},3}
  \,, \nn \\
G_{i, {\rm obs},1} &\equiv \int_0^\pi\! \df \phi \int^{1}_{0}\!\df z \int_{B\mu^2}^{\infty}\!\df s\, \Bigl[Q_i M_{\rm obs} - Q_{i,0} M_{{\rm obs},0} \Theta(z<A)- Q_{i,1}M_{{\rm obs},1} \Theta(1-z<A) \Bigr] 
\nn \\ & \quad
+ \int_0^\pi\! \df \phi \int^{1}_{0} \!\df z \int_{0}^{B\mu^2} \!\df s \Bigl[Q_i(M_{\rm obs}-1) - Q_{i,0}(M_{{\rm obs},0}-1)\Theta(z<A)
\nn \\ & \qquad
 - Q_{i,1}(M_{{\rm obs},1}-1)\Theta(1-z<A)\Bigr]
  \,, \nn \\
G_{i, {\rm obs},2}& \equiv \int_0^\pi\! \df \phi \int^{1}_{0}\! \df z\, \int_{0}^{\infty}\! \df s\, \Bigl[Q_{i,0}M_{{\rm obs},0}\Theta(z<A) + Q_{i,1} M_{{\rm obs},1}\Theta(1-z<A)\Bigr]
  \,, \nn \\
G_{i, {\rm obs},3}&\equiv \int_0^\pi\! \df \phi \int^{1}_{0}\! \df z\, \int_{0}^{B\mu^2}\! \df s\, \Bigl[Q_i - Q_{i,0}\Theta(z<A) - Q_{i,1}\Theta(1-z<A) \Bigr]
\,,\end{align}
where the arguments $s,z,\phi$ are suppressed and $A,B$ are positive real numbers with $A\leq1$. The first term in $G_{i, {\rm obs},1}$ corresponds to the finite part defined in \eq{finitepart}, and the other terms correspond to integrated counterterms. It is straightforward to check that the sum of $G_1$, $G_2$ and $G_3$ is equal to the original one-loop jet function. 

The advantage of the above decomposition is that $G_3$ is observable independent, $G_2$ only depends on the soft limit of the observable (which can be encoded by a few parameters at one-loop order, see \eq{sobs}) and $G_1$ is finite. In \eq{decomp}, $Q_0$ and $Q_1$ denote the soft $z\to 0$ and $z\to 1$ limit of $Q$. Explicitly, 
\begin{align} \label{eq:Qlim}
 Q_{q,0}(s,z,\phi) &=  Q_q(s,z,\phi)|_{z \to 0} = 0\,,
  \nn \\
 Q_{q,1}(s,z,\phi) &=  Q_q(s,z,\phi)|_{z \to 1} = 
\frac{(\mu^2e^{\gamma_E})^{\epsilon}  }{ \sqrt{\pi}\,\Gamma(\tfrac{1}{2}-\epsilon)}  \Bigl(\frac{\nu}{\omega}\Bigr)^{\eta} \frac{2 C_F(\sin\phi)^{-2\eps}}{(1-z)^{1+ \eta+ \epsilon} s^{1+\epsilon}} \,, \nn \\  
  Q_{g,1}(s,z,\phi) &=   \frac{(\mu^2e^{\gamma_E})^{\epsilon}  }{ \sqrt{\pi}\,\Gamma(\tfrac{1}{2}-\epsilon)} \Bigl(\frac{\nu}{\omega}\Bigr)^{\eta}\frac{ C_A(\sin\phi)^{-2\eps}}{(1-z)^{1+ \eta+ \epsilon} s^{1+\epsilon}} =  Q_{g,0}(s,1-z,\phi)\,.
\end{align}

Similarly, $M_{{\rm obs},0}$ and $M_{{\rm obs},1}$ denote the soft $z\to 0$ and $z\to 1$ limit of the measurement $M_\text{obs}$. The soft limit can contain multiple boundary conditions on the phase space, which we account for by writing $M_{{\rm obs},0}$ and $M_{{\rm obs},1}$ as a sum of Heaviside theta functions that constrain the integration over $s$ as a function of $z$. Moreover, they will follow a power-law behavior parametrized by 
\begin{align} \label{eq:sobs}
    M_{\rm obs}(s,z,\phi)|_{z \to 0} &=\Theta(\Phi) \sum_{r}M_{\text{obs}}^{r}=\Theta(\Phi) \sum_r\Theta\Bigl(\frac{c_{0r}^+\, \mu^2}{ z^{\alpha_{0r}^+ }} - s\Bigr) 
    \Theta\Bigl(s- \frac{c_{0r}^- \mu^2}{ z^{\alpha_{0r}^- }}\Bigr)  \, , \\
M_{\rm obs}(s,z,\phi)|_{z \to 1} &=\Theta(\Phi) \sum_{r}M_{\text{obs}}^{r}= \Theta(\Phi) \sum_r\Theta\Bigl(\frac{c_{1r}^+\, \mu^2}{ (1-z)^{\alpha_{1r}^+ }} - s\Bigr) 
    \Theta\Bigl(s- \frac{c_{1r}^- \mu^2}{ (1-z)^{\alpha_{1r}^- }}\Bigr)  \, ,\nn
\end{align}
where the sum on $r$ is over different regions (see \fig{phsp}), and the parameters $c_i$, $\al_i$ depend on the observable, and can depend on $\phi$ as well.\footnote{In general $c_0= c_1$ and $\al_0 = \al_1$, but we will show examples where this is no longer true because the observable depends on the azimuthal angle, which differs by $\pi$ between the two partons.}
We also allow for a constraint $\Phi$ on the azimuthal angle, as will be discussed in \sec{IRS}.
Depending on the observable, each soft boundary condition will therefore follow one out of three distinct behaviors shown in figure \ref{fig:phsp}:  the upper boundary of $R_{1}$ corresponds to $\alpha<0$, the lower boundary of $R_{2}$ to $\alpha=0$, the upper boundary to $\alpha>0$ and $R_{3}$ does not extend into the soft region. Finding $c_{0,1}$ and $\al_{0,1}$ can be nontrivial, and we will discuss a strategy to do so for an involved example in \sec{new}.

\begin{figure}[t!]
\begin{subfigure}{0.31\textwidth}
\includegraphics[width=1\linewidth]{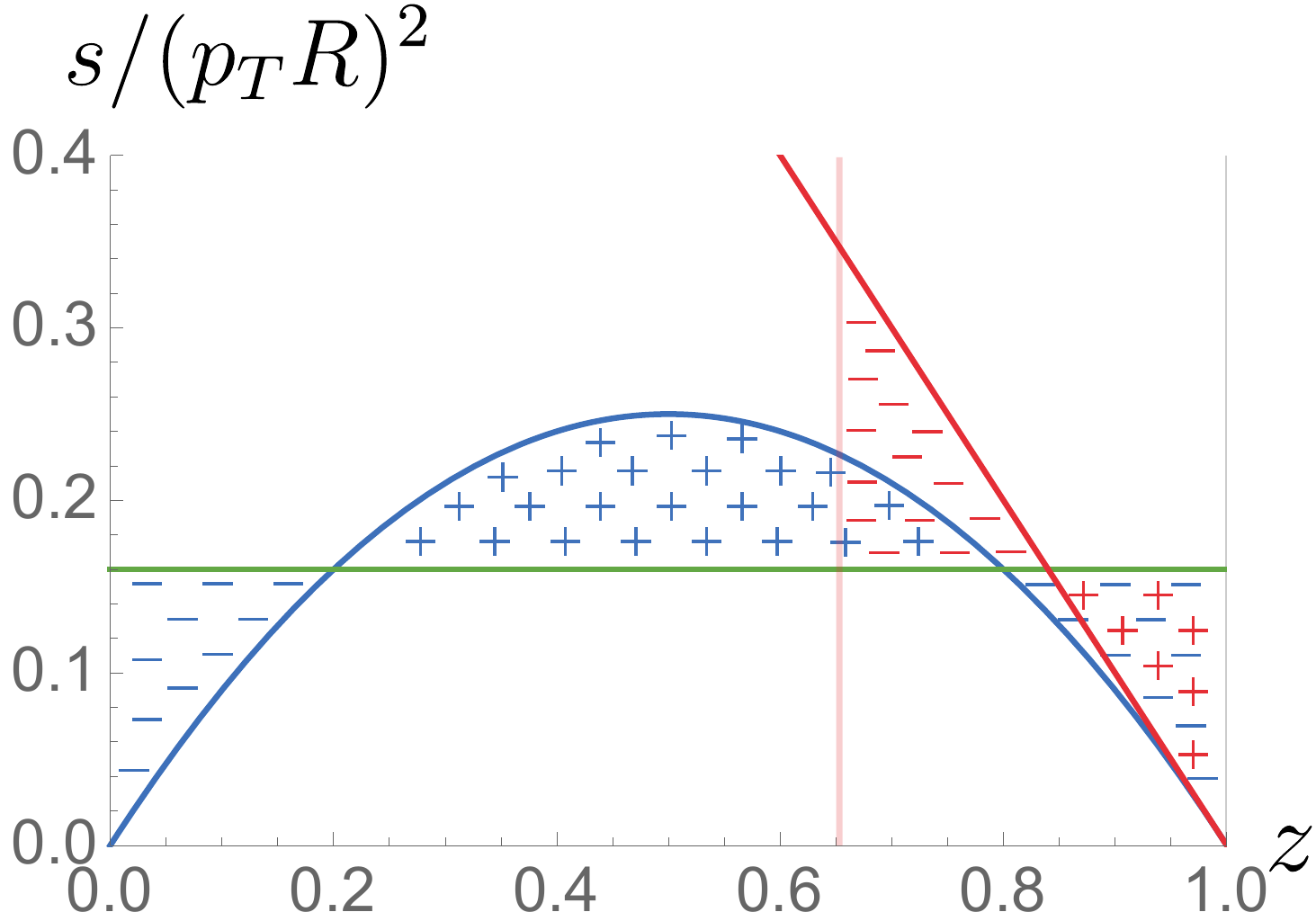} 
\caption{$G_1$: \!Numerical contribution}
\label{fig:kt_g1}
\end{subfigure}\hspace{0.005\textwidth}
\begin{subfigure}{0.31\textwidth}
\includegraphics[width=1\linewidth]{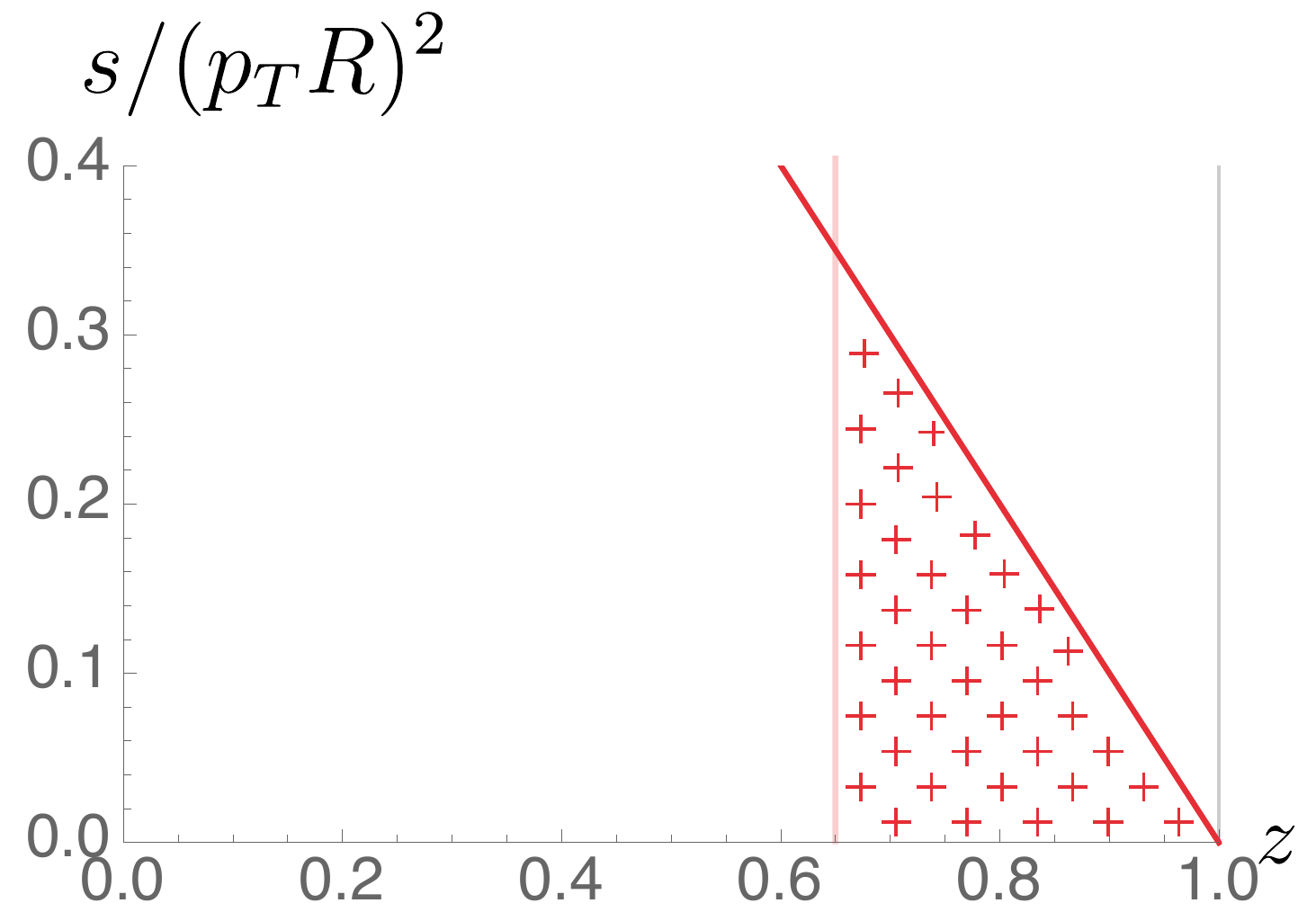}
\caption{$G_2$: Soft counterterm}
\label{fig:kt_g2}
\end{subfigure}\hspace{0.005\textwidth}
\begin{subfigure}{0.31\textwidth}
\includegraphics[width=1\linewidth]{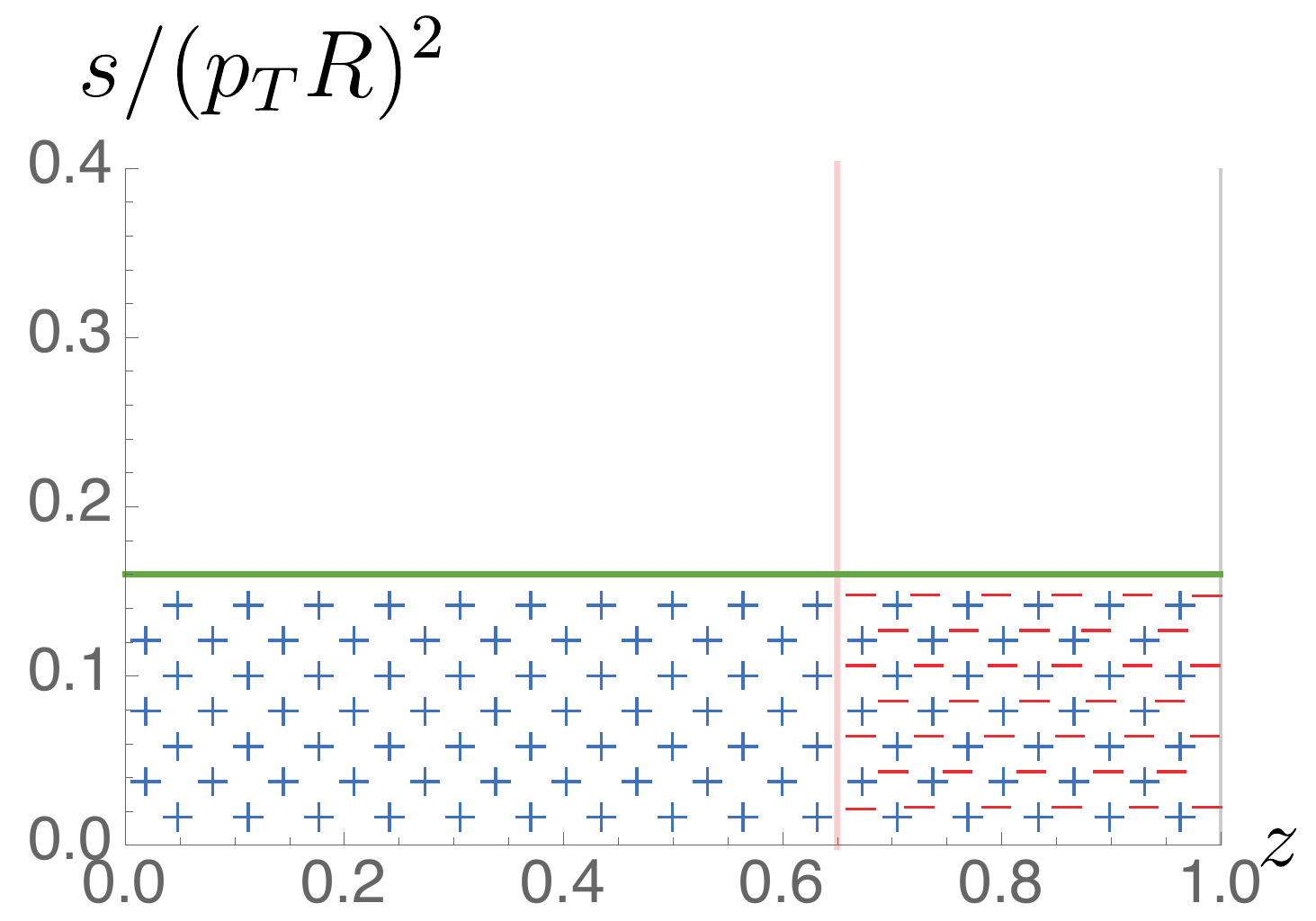}
\caption{$G_3$: Box counterterm}
\label{fig:kt_g3}
\end{subfigure} \\
\caption{A graphical representation of our subtraction scheme in \eq{decomp}. We have only included the soft counterterms for $z\rightarrow 1$ for legibility. Shown are the restriction on the measurement from the observable (blue line), the soft limit of the observable (red line), the box (green line), the cut on $z$ arising from $A$ (pink line). Blue plus (minus) areas correspond to positive (negative) contributions of the full integrand $Q_i M_{\rm obs}$, while red plus (minus) areas correspond to positive (negative) contributions of $Q_{i,1}M_{{\rm obs},1}$.}
\label{fig:kt_example}
\end{figure}

We will now discuss the decomposition in \eq{decomp} in more detail, using the graphical representation in \fig{kt_example} for the $k_T$ algorithm. In order to get a finite $G_1$ in \fig{kt_g1}, we subtracted the collinear singularity and the soft singularities. The collinear singularity is removed by the box, replacing $M_{\rm obs}$ by $M_{\rm obs}-1$ when $s \leq B \mu^2$, such that $M_{\rm obs}(s=0,z,\phi) - 1=0$. The soft singularities get accounted for by subtracting the $z\to0$ and/or $z\to1$ limits of the integrand. Indeed, one can see that in \fig{kt_g1} the blue plusses and red minuses cancel as $z\to1$.
The resulting integral $G_1$ is now finite. For general observables, $G_1$ in \eq{decomp} may be hard to calculate analytically, and one has to resort to numerical integration techniques. In the examples in section \sec{ex}, we will use the \textsc{Cuba} implementation of \Vegas~\cite{cuba} to perform the integrations.
Convergence problems in the numerical integration may arise due to the mismatch of the observable and its soft approximation, which generally  can lead to integrable singularities.  If these problems are severe it can help to find an explicit remapping of the counterterm, which decreases the mismatch between the observable and its soft limit. We present a method for how this can be achieved with a worked through example in \app{CTremap}.

Let us now discuss the integrated counterterms. Due to their simplicity, the counterterms can be calculated analytically, which we discuss for a single region $r$ in the sum in \eq{sobs}. Let us first focus on the soft counterterms, which are contained in $G_2$ shown in \fig{kt_g2}. The soft limits of the integrand $Q_i M_{\text{obs}}$ are given by $Q_{i,0} M_{\text{obs},0}$ and $Q_{i,1} M_{\text{obs},1}$, see \eqs{Qlim}{sobs}. The constants $c_i$ and $\alpha_i$ are user input in our code, see \sec{prog}. For values $\alpha \neq 1$, no rapidity regulator is needed and $\eta$ can be set to 0, leading to the following soft counterterm
\begin{align}
\label{eq:method6}
G_{q,2} &= \frac{2 C_{F}}{\epsilon^2} \frac{e^{\gamma_E\epsilon}  }{ \sqrt{\pi}\,\Gamma(\tfrac{1}{2}-\epsilon)}\int_0^\pi\! \df \phi\, \Theta(\Phi)(\sin\phi)^{-2\eps}    \left[\frac{(c_1^+)^{-\epsilon}}{(1-\alpha_1^+)} A^{-\eps(1-\alpha_1^+)} - \frac{(c_1^-)^{-\epsilon}}{(1-\alpha_1^-)} A^{-\eps(1-\alpha_1^-)} \right] \, , \nn \\
G_{g,2} &= \frac{ C_{A}}{\epsilon^2}  \frac{e^{\gamma_E\epsilon}  }{ \sqrt{\pi}\,\Gamma(\tfrac{1}{2}-\epsilon)} \int_0^\pi\! \df \phi\,  \Theta(\Phi)(\sin\phi)^{-2\eps} \bigg[\frac{(c_0^+)^{-\epsilon}}{(1-\alpha_0^+)} A^{-\eps(1-\alpha_0^+)} - \frac{(c_0^-)^{-\epsilon}}{(1-\alpha_0^-)} A^{-\eps(1-\alpha_0^-)}  \nonumber \\
&\hspace{3.3cm} + \frac{(c_1^+)^{-\epsilon}}{(1-\alpha_1^+)} A^{-\eps(1-\alpha_1^+)} - \frac{(c_1^-)^{-\epsilon}}{(1-\alpha_1^-)} A^{-\eps(1-\alpha_1^-)} \bigg].
 \end{align}
For $\alpha=1$ one needs a rapidity regulator and the corresponding expression is given in \app{rap}. The box counterterm $G_3$ in \fig{kt_g3} is given by 
\begin{align}
\label{eq:method6b}
G_{q,3}&=C_{F} \,I(\phi^+,\phi^-;\eps)\,\frac{e^{\gamma_E\epsilon}B^{-\eps}  }{ \sqrt{\pi}\,\Gamma(\tfrac{1}{2}-\epsilon)}
\left(\frac{(4-\eps) (1-\eps)\Gamma^2[1-\eps] }{2\Gamma[2-2\eps]}-2A^{-\eps} \right), \\
G_{g,3}&=\,I(\phi^+,\phi^-;\eps)\,\frac{e^{\gamma_E\epsilon}B^{-\eps}  }{ \sqrt{\pi}\,\Gamma(\tfrac{1}{2}-\epsilon)}\left(-\Big(\frac{3}{2} C_{A} (3 \eps-4)+2 \eps\  n_{f}  T_{R}\Big)\frac{(1-\eps) \Gamma^2[1-\eps]}{(3-2\eps)\Gamma[2-2\eps]}
-2 C_{A} A^{-\eps}\right).\nn 
\end{align}
The integral over $\phi$ has been carried out for $\Theta(\Phi)=\Theta(\phi^+-\phi)\Theta(\phi-\phi^-)$ leading to the function  
\begin{equation}
I(a,b;\eps)=\int_a^b \! \df \phi\, \sin^{-2\eps}\phi\,.
\end{equation}
The evaluation of this integral and its expansion to order $\eps^2$ is presented in \app{azimuth}. 

The chosen subtraction bears fruit in the simplicity of the integrated counterterms. The corresponding Laurent series in $\eps$ can be expressed solely in terms of the Riemann zeta function at integer values, given that only pure Gamma functions appear. From an analytic point of view, the potentially more complicated pieces are instead captured in the finite part, which depends on the details of the observable and can be calculated numerically to arbitrary high order in $\eps$. Notice  that the soft counterterm $G_{i,2}$ can give rise to more complicated integrals if the coefficients $c_{i}^\pm$ depend on the azimuthal angle $\phi$. One may be able to carry out this integral analytically in certain cases, but this can certainly not be done in general. This is not a problem, because one can expand in $\eps$ and $\eta$ before integrating over $\phi$.

\subsection{Delta and theta functions}
\label{sec:dfs}

In our subtraction scheme we assume that the observables restrict the integration to certain regions of phase space via Heaviside theta functions. However, many observables $\cO$ are naturally expressed in terms of Dirac delta functions, requiring one to rewrite it using 
\begin{align}
\label{eq:df1}
\de[\cO - f(s,z,\phi)] = \pm \frac{\df}{\df \cO} \Theta[\pm(\cO-f(s,z,\phi))]
\,,\end{align}
where $f$ is a function of the kinematics of the collinear splitting, and possibly external parameters.
The sign $\pm$ should be chosen such that the theta function does not vanish at tree-level, which ensures that the poles are included in the one-loop jet function. For example, if $\cO \geq 0$ and at tree-level $\cO=0$, one needs to choose the plus sign in \eq{df1}.

In perturbative QCD one often works with the following convention for the Dirac delta function, 
\begin{align}
\label{eq:physicsdelta}
g(0) &= \int^{c}_{0}\! \df x \ g(x)\delta(x) \,\qquad \text{for}\ c>0\,.
\end{align}
This differs from the definition given in standard math literature
\begin{align}
g(0) &= \int^{c}_{b}\! \df x \ g(x)\delta(x) \,\qquad \text{for}\ c>0>b\,,
\end{align}
where the lower boundary $b$ must be strictly less than zero.
If the delta function that encodes the measurement satisfies \eq{physicsdelta}, this has implications for the definition of the Heaviside function on the right-hand side of \eq{df1}. In particular, one must demand then that $\Theta(0)=0$. To see this, consider a function $g(x)$ with $0\leq x\leq 1$. From
\begin{align}
\label{method:df3}
g(0) &= \int^{1}_{0}\! \df x \ g(x)\delta(x) = \int^{1}_{0} \df x \ g(x)\, \frac{\df}{\df x}\Theta(x) 
= [g(x)\Theta(x)]^{1}_{0} - \int^{1}_{0}\! \df x\, \frac{\df}{\df x}g(x)
\nn \\ 
&= g(1)\Theta(1) - g(0)\Theta(0) - (g(1) - g(0))
= g(0)(1-\Theta(0))
\,,\end{align}
we conclude that $\Theta(0) =0$. While this is not of much concern when a theta function is integrated over, there are situations where it must be taken into account. 
As an example, the jet shape calculation involves a jet function describing the energy fraction $z$ inside a cone, see \sec{pedro}. Switching to a cumulant variable for $z$, we need to choose $\de(z - \dots) = - \df/\df z [-(z - \dots)]$, because $0 \leq z \leq 1$ and $z=1$ at tree-level. If we now want to calculate the average momentum fraction from the cumulant tree-level result
\begin{align}
  \int_0^1 \df z\, z\, \de(z-1) =  -\int_0^1 \df z\, z\, \frac{\df}{\df z}\, \theta(1-z) = - z \theta(1-z)|^1_0
  + \int_0^1 \df z\, \theta(1-z)
   = 1 - \theta(0) =1
\,,\end{align}
we have to take $\theta(0)=0$ to find agreement with the direct evaluation using the delta function.

\subsection{Infrared safety and limitations on the observable} 
\label{sec:IRS}

\begin{figure}[t]
{\centering
\begin{subfigure}{0.32\textwidth}
\includegraphics[width=1\linewidth]{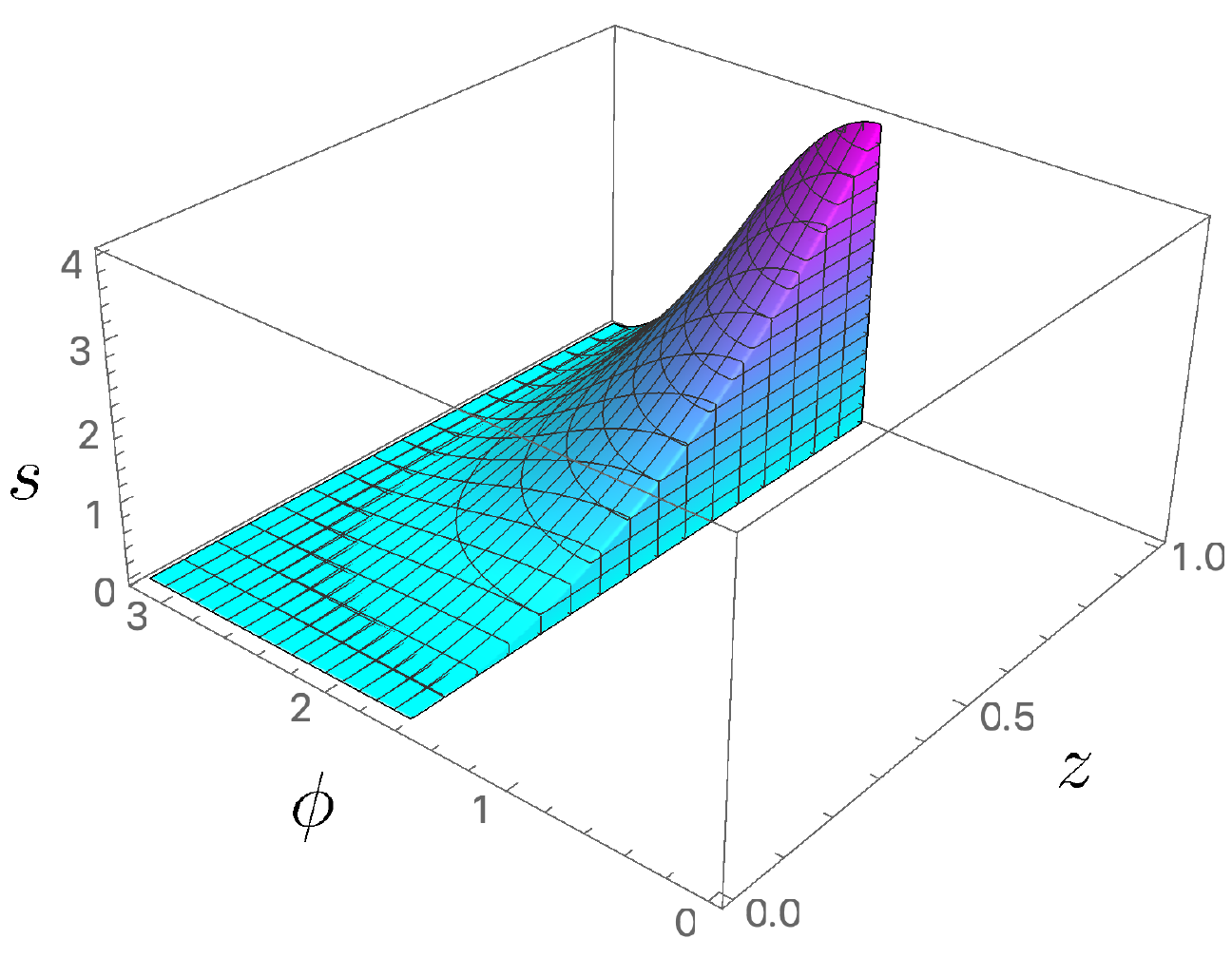} 
\caption{}
\label{fig:IRunsafeslice_a}
\end{subfigure} \qquad
\begin{subfigure}{0.32\textwidth}
\includegraphics[width=1\linewidth]{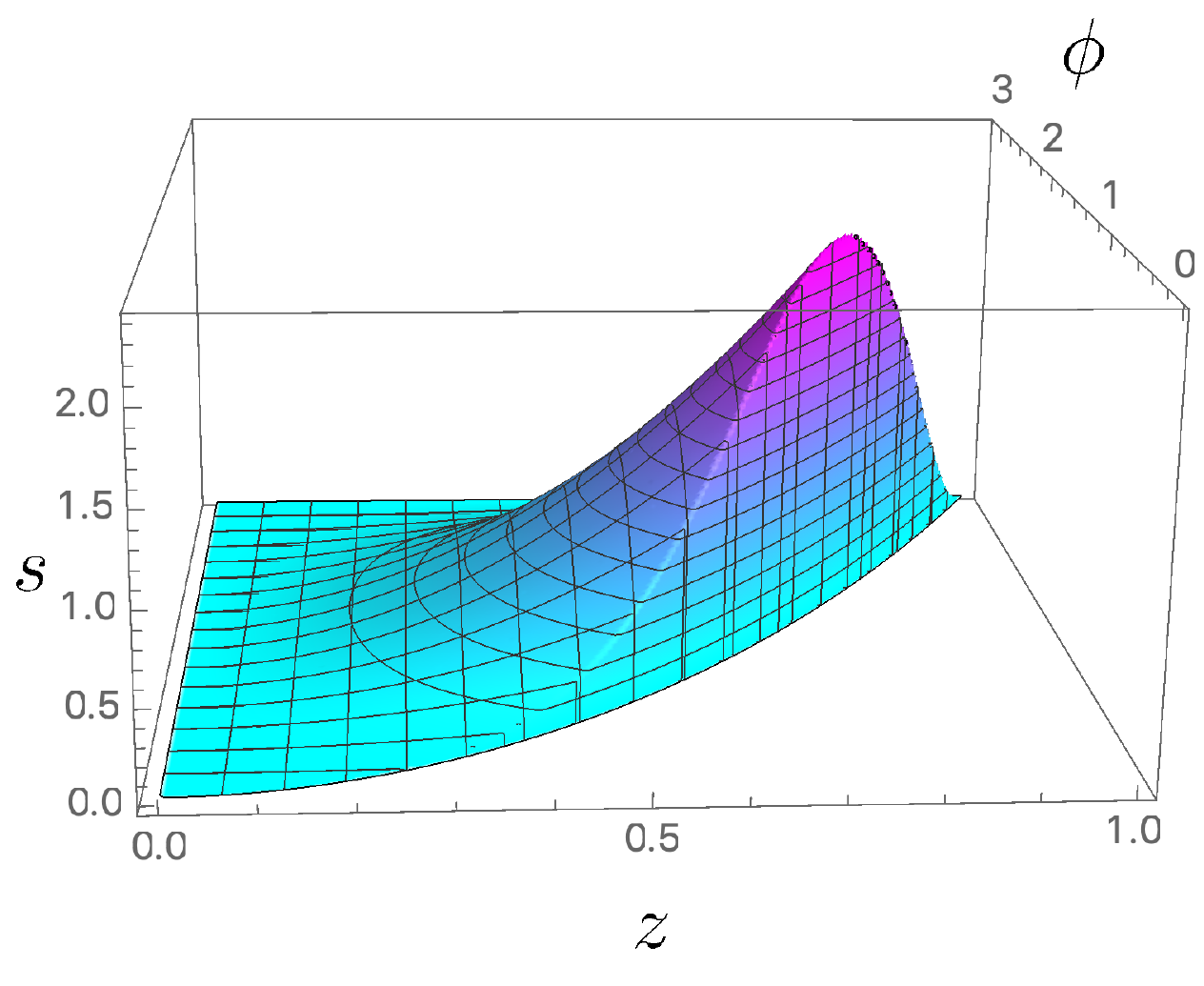}
\caption{}
\label{fig:IRunsafeslice_b}
\end{subfigure}\\}
\caption{IR unsafe observables that our code (a) can and (b) can't handle.}
\label{fig:IRunsafeslice} 
\end{figure}

While so far our discussion was mostly based on the $s$-$z$ plane, there are observables which depend also on the azimuthal angle $\phi$. 
The integration domain is then parametrized by coordinates $(s,z,\phi)$ and IR safety requires the full $s=0$ plane to be included or excluded by the observable, i.e.~the set of points 
\begin{align}
\{(s,z,\phi)|s=0\,,0\le z\le 1\,,0\le\phi\le\pi \}\,.
\end{align} 
However, our method allows for a special class of IR-unsafe observables, where only subdomains of the collinear plane with the azimuthal angle bounded between constant values are included/excluded by the observable, i.e.
\begin{align}
\{(s,z,\phi)|s=0\,, 0\le z\le1\,, \phi^-\le\phi\le\phi^+\}\,,
\end{align}
with $0\le\phi^-<\phi^+\le \pi$. This is illustrated in \fig{IRunsafeslice_a}. An IR-unsafe observable which is not of this form, and currently not supported by \gojet, is illustrated in \fig{IRunsafeslice_b}. Here $\phi^\pm$ vary as functions of $z$ across the collinear plane in such a way that not the full $s=0$ plane is included in the integration domain. For $s>0$ the bounds on $\phi$ can depend on $z$. \gojet can also handle IR-unsafe observables that include just $z=0$ and/or $z=1$ of the $s=0$ plane, which only require soft counterterms.

\subsection{Example: Angularities with the Winner-Take-All axis}
\label{sec:exang}

We will now illustrate our scheme by considering the family of $e^+e^-$ event shapes called angularities~\cite{Berger:2003iw}
\begin{align} \label{eq:ea_def}
  e_b &\equiv \frac{1}{Q} \sum_i E_i (\sin \theta_i)^{1-b} (1-|\cos \theta_i|)^{b} \stackrel{\theta_i \ll 1}{\approx} \frac{2^{-b}}{Q} \sum_i E_i \theta_i^{b+1}
\,,\end{align}
parametrized by $b$\footnote{Our $b$ is related to the parameter $a$ in ref.~\cite{Berger:2003iw} by $b=1-a$.}.
Here $Q$ is the center-of-mass energy, and the sum runs over all particles $i$ in the final state with energy $E_{i}$ and angle $\theta_{i}$ with respect to some axis.  The final expression is only valid in the small-angle limit, which is appropriate for the jet function calculation, highlighting that $e_b$ probes the angular distribution with exponent $1+b>0$.  While angles were originally taken with respect to the thrust axis, we will here use the Winner-Take-All axis~\cite{Larkoski:2014uqa}. For the one-loop jet function this axis is simply along the most energetic particle in the jet, so the only non-zero contribution in the sum on $i$ in \eq{ea_def} comes from the least energetic particle, with $\theta_i$ the angle between the two partons in the jet. Noting that $s = 2p_1 \cdot p_2 = \tfrac12 z(1-z)(1- \cos \theta) Q^2 \approx \tfrac14 z(1-z)\theta^2 Q^2$,  we obtain the following measurement function for a cut on the angularity $e_b \leq e_b^c$,
\begin{align}
  M_b(s,z) = \Theta\Bigl[z (1 - z) \, Q^2 \Bigl(\frac{e_b^c}{\min[z, 1 - z]}\Bigr)^{2/(b+1)}  - s\Bigr].
\end{align}
For angularity exponent $b<1$, the observable is unbounded from above, similar to the top curve of region 2 in \fig{phsp}. In the notation of \eq{sobs}, we see that the soft limit of the observable is characterized by $c_0= c_1= Q^2 (e_b^c)^{2/(b+1)}/\mu^2$ and $\al_0=\al_1 = 2/(1+b) - 1$. The one-loop contribution to the jet function is obtained by plugging in these these constants in \eqs{method6}{method6b} to calculate $G_2$, performing the integration over $s$ and $z$ for $G_1$, and adding these contributions to the box $G_3$. Performing the integration over $s$ analytically and the integration over $z$ numerically for $b=2$, we obtain 
\begin{align}
\label{eq:method11}
\mathcal{J}^{(1)}_{q,e_2} = \frac{\alpha_s C_F}{2\pi} \biggl(\frac{ \mu^2}{Q^2 (e_2^c)^{2/3}}\biggr)^{\eps} \Bigl(\frac{3}{2 \epsilon^2} + \frac{3}{2 \epsilon}-1.909961286856877\Bigr),
\end{align}
where we used $\mu = Q (e_2^c)^{1/3}$ to calculate the constant contribution and reinstated the logarithmic behaviour afterwards. Our result agrees with the expression in refs.~\cite{Larkoski:2014uqa,Procura:2018zpn} up to order $10^{-11}$.\footnote{Refs.~\cite{Larkoski:2014uqa,Procura:2018zpn} both use $\bt = 1+b$ instead of $b$, and ref.~\cite{Larkoski:2014uqa} also removes the $2^{-b}$ from the definition in \eq{ea_def} and takes $Q$ to be the jet energy.}  For $b=0$ the rapidity regulator is required. In that case we find 
\begin{align}
\label{eq:method12}
\mathcal{J}^{(1)}_{q,e_0} = \frac{\alpha_s C_F}{2\pi} \biggl(\frac{2\nu}{Q}\biggr)^{\eta} \biggl(\frac{ \mu^2}{Q^2 (e_0^c)^{2}}\biggr)^{\eps} \Bigl(\frac{2}{ \epsilon\eta} + \frac{3 - 4 \log{2}}{ 2 \epsilon} -1.8693096781349734 \Bigr),
\end{align}
in agreement with ref.~\cite{Larkoski:2014uqa}.

\section{\gojet Program}
\label{sec:prog}

The \gojet \Mathematica-package automatically performs the subtraction, given the observable and its soft limit (see \eq{sobs}) as input. One can either let \Mathematica perform the numerical integration or choose to export the integrand. The latter feature may be useful if \texttt{NIntegrate} either has difficulty converging or is not fast enough. In such cases it can be advantageous to use algorithms such as \Vegas, that are faster due to their implementation in C++ or Fortran. A general overview of the various functions included in the package is given in \sec{functions}. A detailed description of their input is given in \sec{input}, with a worked-out example in \sec{example}. 
 
\subsection{Functions}
\label{sec:functions}
There are a total of 12 different functions, listed in \sec{input}, which the user can access. As indicated by their names half of these are for calculating gluon jet functions while the other half are for calculating quark jet functions. Restricting to the former, \texttt{PolesGluon} returns the pole terms in $\eps$ and $\eta$ for the gluon jet function and \texttt{GluonJet} returns the integrand of the finite terms, by which we here refer to the $\eps^0 \eta^0$-term. In addition, \texttt{GluonJetN} performs the numerical integration over the cube $0 \leq s,z,\phi \leq 1$ of this integrand. This integration domain is the result of mapping $s \to s/(1-s)$ and $\phi \to \pi \phi$, which also stabilizes the integration over $s$. Note that \texttt{GluonJet} also contains the $\eps^0 \eta^0$-pieces of the counterterms $G_2$ and $G_3$, which are already integrated over analytically. For the convenience of the user these pieces are simply added in integrated form since they are not altered by the  trivial numerical integration over the unit cube. 

Let us now discuss the arguments of the functions in general terms. The first arguments encode the measurement $\texttt{O}$ and its soft limit $\texttt{O}_0$ and $\texttt{O}_1$ corresponding to the limits $z\to 0$ and $z \to 1$, respectively. The observable should generally be IR safe, with some exceptions discussed in \sec{IRS}. Furthermore, we require certain restrictions on the form of the soft limits. Specifically, it is not possible to restrict the $\phi$-integration boundaries via $\texttt{O}_0$ and $\texttt{O}_1$, whose format is fixed. It is however possible to apply $s,z$-independent constraints on the boundaries of the  $\phi$-integration through the separate argument $\Phi$, which are the same for the finite part as well as the counterterms. 

The next set of arguments specify the regularization and IR scheme: the need of a rapidity regulator or collinear regulator is controlled by the switches \texttt{rr} and \texttt{box}, respectively. The explicit cut for the soft limits and box is specified by \texttt{A} and \texttt{B} (see \eq{decomp}). The independence of the final result on these parameters provides a useful cross-check for the calculation. A specific choice of these parameters can also be used to improve the convergence of the numerical integration. For the gluon jet function, the number of quark flavors is specified through the argument \texttt{nf}. The number of colors has been fixed to three, but the full dependence on the Casimirs can be easily reconstructed from the answer. The final set of arguments enables the user to specify the integration method or output format for the integrand.

Finally, we also allow for more complicated observables, where the phase-space restriction due to the measurement breaks up into more than one region. The corresponding functions have ``\texttt{Regions}" appended to their name, and contain additional arguments specifying possible dependence on external parameters in the regions.

\subsection{Input format}
\label{sec:input}
Here we specify the syntax of each of the functions: \\

\noindent \texttt{GluonJet[O, \!$\text{O}_0$,  \!$\text{O}_1$, \!$\Phi$, \!rr, \!box, \!A, \!B, \!s, \!z, \!$\phi$, \!nf, \!format, \!file] \vspace{0.1cm}}

\noindent \texttt{GluonJetRegions[R, \!O, \!$\text{R}_0$, \!$\text{O}_0$, \!$\text{R}_1$, \!$\text{O}_1$, \!$\Phi$, \!rr, \!box, \!A, \!B, \!s, \!z, $\phi$, \!nf, \!format, \!file]} 

\noindent \texttt{GluonJetN[O, \!$\text{O}_0$,  \!$\text{O}_1$, \!$\Phi$, \!rr, \!box, \!A, \!B, \!s, \!z, \!$\phi$, \!nf, \!method]\vspace{0.1cm}}

\noindent \texttt{GluonJetRegionsN[R, \!O, \!$\text{R}_0$, \!$\text{O}_0$, \!$\text{R}_1$, \!$\text{O}_1$, \!$\Phi$, \!rr, \!box, \!A, \!B, \!s, \!z, $\phi$, \!nf, \!method] \vspace{0.1cm}}

\noindent \texttt{PolesGluon[$\text{O}_0$, \!O$_1$, \!$\Phi$, \!rr, \!box, \!A, \!B, \!$\phi$, \!nf] \vspace{0.1cm}} 

\noindent \texttt{PolesGluonRegions[$\text{O}_0$, \!$\text{O}_1$, \!$\Phi$, \!rr, \!box, \!A, \!B,  \!$\phi$, nf] \vspace{0.1cm}} \\

\noindent \texttt{QuarkJet[O, \!$\text{O}_0$,  \!$\text{O}_1$, \!$\Phi$, \!rr, \!box, \!A, \!B, \!s, \!z, \!$\phi$, \!format, \!file] \vspace{0.1cm}}

\noindent \texttt{QuarkJetRegions[R, \!O, \!$\text{R}_0$, \!$\text{O}_0$, \!$\text{R}_1$, \!$\text{O}_1$, \!$\Phi$, \!rr, \!box, \!A, \!B, \!s, \!z, $\phi$, \!format, \!file] \vspace{0.1cm}}

\noindent \texttt{QuarkJetN[O, \!$\text{O}_0$, \!$\text{O}_1$, \!$\Phi$, \!rr, \!box, \!A, \!B, \!s, \!z, \!$\phi$, \!method]\vspace{0.1cm}}

\noindent \texttt{QuarkJetRegionsN[R, \!O, \!$\text{R}_0$, \!$\text{O}_0$, \!$\text{R}_1$, \!$\text{O}_1$, \!$\Phi$, \!rr, \!box, \!A, \!B, \!s, \!z, $\phi$, \!method]\vspace{0.1cm}}

\noindent \texttt{PolesQuark[$\text{O}_0$, \!O$_1$, \!$\Phi$, \!rr, \!box, \!A, \!B, \!$\phi$] \vspace{0.1cm} }

\noindent \texttt{PolesQuarkRegions[$\text{O}_0$,  \!$\text{O}_1$, \!$\Phi$, \!rr, \!box, \!A, \!B, \!$\phi$] \vspace{0.1cm}} 
\\

\noindent The variables used to describe the input are:
 \begin{itemize}
 \item \texttt{O}: The list of argument(s) of the Heaviside theta function encoding the bounds imposed by the measurement. More specifically, $\texttt{O}$ contains the arguments of the Heaviside theta functions $M_{\text{obs}}$ in \eq{method1}. For the case of a single region, the elements of the list correspond to the arguments of Heaviside theta functions, whose product constrain the region. In the case of multiple regions, $\texttt{O}$ is a list of lists. The entries of the outer list correspond to the different regions, each entry is again a list of constraints containing the arguments of the Heaviside theta functions $M_{\text{obs}}^{r}$ constraining the particular region. This allows the user to implement arbitrary sums of products of Heaviside theta functions. 
 
 \item $\texttt{R}_1$ ($\texttt{R}_0$): List of lists which contain arguments of Heaviside theta functions which  depend \emph{only} on external parameters for each region in the limit $z\rightarrow 1$ ($z\rightarrow 0$). The length of this list is therefore equal to the number of soft regions that emerge in the soft limit. Regions that do not depend on external parameters need $\{1\}$ as input in their respective position in the list. The number of soft regions can be less than the number of regions, but should match with the lists for $\texttt{O}_0$ and $\texttt{O}_1$ below. In particular, regions may merge or disappear in the soft limit. $\texttt{R}_1$ ($\texttt{R}_0$) can also be used in cases with just one region where there is dependence on external parameters in the soft limits.
 
 \item $\texttt{O}_1$ ($\texttt{O}_0$): List $\{\{ c_{1}^-, \alpha_1^- \}, \{ c_1^+,\alpha_1^+ \} \}$ describing the lower and upper boundary of the region in the limit where $z\rightarrow 1$ (and similarly for $z\rightarrow 0$), see \eq{sobs}. If there is no lower boundary, $c_1^-$ is just 0. When considering multiple regions, $\texttt{O}_1$ ($\texttt{O}_0$) is a list of lists where each region has an upper and a lower boundary of the aforementioned format. 
 \item $\Phi$: List of arguments of the Heaviside theta functions that impose constraints on the azimuthal angle $\phi$, i.e., the input $\{\phi^+-\phi,\phi-\phi^-\}$ will constrain $\phi^-<\phi<\phi^+$. In the case of multiple regions that contain collinear and/or soft divergences we require the range on $\phi$ to be the same for all regions. (Arbitrary constraints on $\phi$ can of course be encoded in \texttt{O}; but these are not allowed to survive singular limits; that is the they should match the boundaries imposed by $\Phi$ in these limits; see \sec{IRS} for more details.)
  
 \item \texttt{rr}: Boolean variable specifying whether a rapidity regulator should be included, which we implemented as 
\begin{align}
\label{eq: rapreg}
\left( 2(1-z) z ) \right)^{-\eta}
 \end{align} 
 This corresponds to the more conventional factor $\left(\nu/ ( (1-z) z \ \omega) \right)^\eta$, for the scale choice $\nu= \tfrac12 \omega$. The user can always reconstruct the full dependence on the scale $\nu$ a posteriori, given the knowledge of the $1/\eta$ pole.
  
 \item \texttt{box}: Boolean controlling whether a box is needed to handle the collinear divergence. It should be included when the region of phase space includes $s=0$ and not otherwise (in line with the restrictions outlined in \sec{IRS}). 
\item \texttt{A}: Real number specifying the region where the soft counterterms are subtracted. Explicitly, the $z\rightarrow 0$ ($z  \rightarrow 1$) counterterms are subtracted in the phase-space region where $z<\texttt{A}$ ($1-z<\texttt{A}$), and therefore $0< \texttt{A} \leq 1$. 
 \item \texttt{B}: Postive real number specifying the size of the box. 
 \item \texttt{s}: Variable used to describe the invariant mass of the parton that initiates the jet. In the code we have made this variable dimensionless by rescaling with the renormalisation scale $\mu^2$, i.e., $\texttt{s}=\frac{s}{\mu^2}$. 
 \item \texttt{z}: Variable encoding the momentum fraction $z$ of one of the partons in the collinear splitting.
 \item $\phi$: Variable corresponding to the azimuthal angle of the collinear splitting.
 \item \texttt{nf}: Variable specifying the number of (massless) quark flavors. This variable does not need to be set to an integer, but can be left in symbolic form.
 \item \texttt{format}: String specifying the output form of this function. One can choose between ``Mathematica", ``Fortran" and ``C".
 \item \texttt{file}: String with the filename to which the integrand will be exported.  For an empty string the integrand will be printed to the screen.  
 \item \texttt{method}: This string can specify which method \texttt{NIntegrate} uses in \Mathematica, and we refer the reader to the \Mathematica documentation for the available options. For an empty string the default method of \texttt{NIntegrate} will be used. 
 \end{itemize}
 
\subsection{Example: $k_T$ clustering algorithms} 
\label{sec:example}

To illustrate the use of our code we now calculate the jet function for the family of $k_T$ clustering algorithms. At one-loop order, where there are at most two particles in the final state, they are clustered into a single jet if the angle between them is less than the jet radius parameter $R$, which for the case of an $e^+e^-$ collider corresponds to a single region\footnote{The corresponding result for $pp$ collisions can be obtained by simply replacing the jet energy $E$ by the jet transverse momentum $p_T$, and $R$ then corresponds to a distance in $(\eta,\phi)$ instead of an angle.\label{footnote}}
\begin{align} \label{eq:sobskt}
s \leq z(1-z) E^2 R^2\,,
\end{align}
where $E$ is the jet energy.
The $z\to 0$ and $z \to 1$ limits of \eq{sobskt} are described by 
\begin{align}
z \to 0: & \quad s = z E^2 R^2 \hspace{1.355cm} \longrightarrow\hspace{0.4cm} c_0^+ = E^2R^2/\mu^2, \alpha_0^+ = -1, \nn \\
z \to 1: & \quad s = (1 - z) E^2 R^2 \hspace{0.4cm} \longrightarrow\hspace{0.4cm}  c_1^+ = E^2R^2/\mu^2, \alpha_1^+ = -1.
\end{align}
These are no lower constraints, i.e.~$c_i^- = 0$. Calculating this observable requires a box since the $s=0$ line is inside the domain of integration. Since $\al_i \neq 1$, a rapidity regulator is not needed. 
The constraint in \eq{sobskt} due to the measurement does not depend on $\phi$, and so we take $\Phi=\{\}$. 

We now calculate the quark jet function. As \eq{sobs} is a relatively simple expression, for which the jet function can be easily calculated analytically, we will use \Mathematica to perform the numerical integration over the subtracted integral by using \texttt{QuarkJetN} with the the `LocalAdaptive' integration method. In the following we set $\mu = ER$ for simplicity. Note how this, since the variable \texttt{s} corresponds to $\frac{s}{\mu^2}$, cancels the factor $E^2R^2$ in the obsevable.
\begin{mma}
  \In \textbf{O = z(1 $-$ z) $-$ s}; \\
\hspace{0.79cm} \textbf{O$_\textbf{0}$ = \{\{\hspace{0.01cm}0\hspace{0.01cm},\hspace{0.02cm}0\hspace{0.01cm}\},\{1,$-$1\}\}; } \\
\indent \hspace{0.79cm}  \textbf{O$_\textbf{1}$ = \{\{\hspace{0.01cm}0\hspace{0.01cm},\hspace{0.02cm}0\hspace{0.01cm}\},\{1,$-$1\}\}; }  \\
\indent\hspace{0.79cm} \textbf{method =``LocalAdaptive";} \\
\indent\hspace{0.79cm} \textbf{box = True;} \\
\indent\hspace{0.79cm} \textbf{rr = False;} \\
\indent\hspace{0.79cm} \textbf{A=0.6;}\\
\indent\hspace{0.79cm} \textbf{B=20;} 
\end{mma}
\begin{mma}
  \In \textbf{QuarkJetN}[\textbf{O},\textbf{O}_0, \textbf{O}_1, \{\}, \textbf{rr}, \textbf{box}, \textbf{A},\textbf{B}, \textbf{s}, \textbf{z}, \textbf{phi}, \textbf{method}] \\
  \Out -1.2029367022'\\
  \In \textbf{PolesQuark}[\textbf{O}_0,\textbf{O}_1, \{\}, \textbf{rr}, \textbf{box}, \textbf{A},\textbf{B}, \textbf{phi}] \\
  \Out \frac{4}{3 \eps^2} + \frac{2}{\eps}\\
\end{mma}
From this answer it is straight forward to reconstruct that the full color-dependence of the regulated one-loop quark jet function is given by:
\begin{align}
\label{eq: ktres}
\mathcal{J}_{q}^{k_{T}} = C_{F}\bigg(\frac{1}{\eps^2}+\frac{3}{2 \eps} -0.9022033008\bigg).
\end{align}
The poles match exactly with the result by \cite{Ellis:2010rwa} and the finite term agrees up to order $10^{-6}$. Similar agreement is found for the gluon jet function:
\begin{align}
\label{eq: ktresgluonjet}
\mathcal{J}_{g}^{k_{T}} = C_{A} \left(0.0422426\, +\frac{1}{\eps^2}+\frac{11}{6 \eps}\right) - n_{f} T_{R} \left(\frac{2}{3 \eps}+2.55555\right).
\end{align}
The accompanying \Mathematica notebook contains several hands-on examples to further illustrate the use of the different functions.

\section{Applications}
\label{sec:ex}

To validate the method and corresponding code, the jet functions for several known examples have been checked. Some of these were used throughout the paper to explain our approach, namely the $k_{T}$ family of clustering algorithms (\sec{example}), and angularities with respect to the WTA axis (\sec{exang}). In addition, we  provide results in \sec{cone} for the cone algorithm and in \sec{pedro} for the jet shape. The latter is more challenging due to its azimuthal-angular dependence, which arises because the jet axis is along the total jet momentum and thus sensitive to recoil of soft radiation. In \sec{new} we present, for the first time, the one-loop jet functions for angularities with respect to the thrust axis, taking into account recoil. Although for $b>0$ this recoil is formally power-suppressed, it can be numerically large~\cite{Budhraja:2019mcz}.

\subsection{Cone jet}
\label{sec:cone}

At one-loop order, the condition that both partons are within a cone jet in an $e^+e^-$ collisions is that their angle with the jet axis is less than $R$ (for $pp$${}^\text{\ref{footnote}}$). Since the jet axis is along the total jet momentum, one simply needs to consider the angle with the parton that initiates the jet, leading to the following condition
\begin{align}
\label{eq: coneobs}
s\leq E^2 R^2 \min\Bigl[\frac{1-z}{z},\frac{z}{1-z}\Bigr].
\end{align}
As we focus on the finite term in the jet function, we fix $\mu=E R$ finding 
\begin{align}
\label{eq: conejet}
\mathcal{J}_{q}^{\text{Cone}}&=C_{F} \Bigl(1.46711\, +\frac{1}{\eps^2}+\frac{3}{2 \eps}\Bigr), \nn \\
\mathcal{J}_{g}^{\text{Cone}}&=C_{A} \Bigl(2.23477\, +\frac{1}{\eps^2}+\frac{11}{6 \eps}\Bigr) - n_{f} T_{R}\Bigl(\frac{2}{3 \eps}+2.20197\Bigr),
\end{align}
which agrees up to order $10^{-6}$ with ref.~\cite{Ellis:2010rwa}.

\subsection{Angularities with recoil}
\label{sec:new}
In this section we determine, for the first time, the one-loop angularity jet function that includes the recoil of the thrust axis due soft radiation. While this recoil is power-suppressed for $b>0$, ref.~\cite{Budhraja:2019mcz} noted that it has a numerically large effect and presented a factorization framework to include it. The one-loop jet function we calculate here will start to contribute at NLL$'$ accuracy. This should be contrasted with the calculation in \sec{exang}, where we considered the angularity with respect to the WTA axis. To clearly distinguish these two cases in the notation, we will use $\tau_n$ instead of $e_b$, where $n$ refers to the thrust axis.

\begin{figure}[t]
\centering
\includegraphics[width=0.7\textwidth]{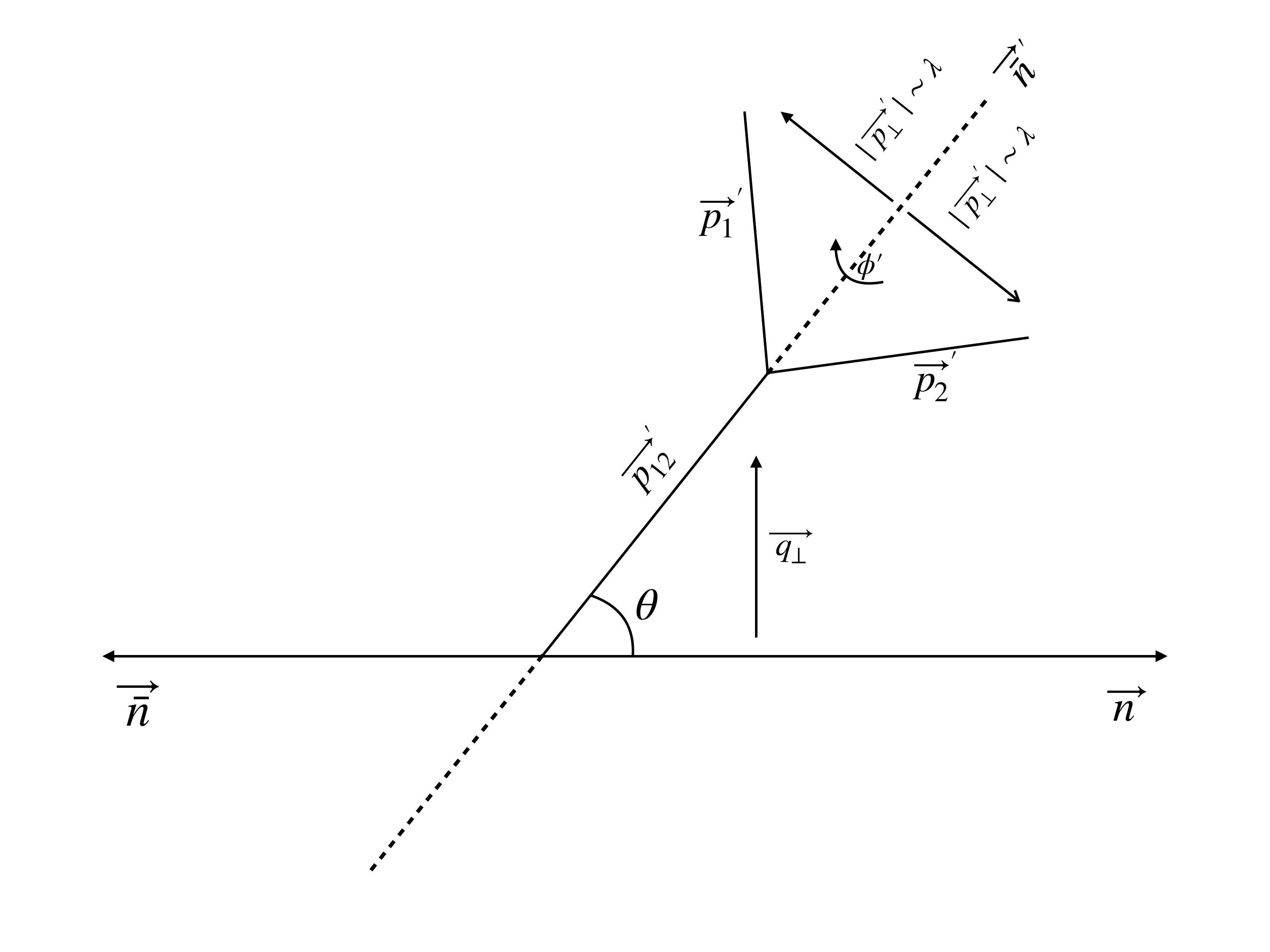}
\caption{The setup of our calculation. The recoil is quantified by $\theta$.}
\label{fig:recang} 
\end{figure}
The setup underpinning our calculation is illustrated in \fig{recang}. Here $\theta$ is the angle between the thrust axis $\vec n$ and the direction $\vec n'$ of the initial collinear parton due to the recoil from soft radiation, which is treated as an external parameter in our calculation. The momenta of the two massless partons in the jet are denoted by $\vec{p_{1}}$ and $\vec{p_{2}}$, where we use (un)primed coordinates to denote light-cone components with respect to the $\vec n'$ ($\vec n$) direction. Explicitly,
\begin{align}
\label{eq:4.3.1}
p_{1}'^{\mu}&= zQ\,\frac{n'^{\mu}}{2} + \frac{(1-z)s}{Q}\,\frac{\bar{n}'^{\mu}}{2} + p_{1\perp}'^{\mu}
\equiv p_1'^-\frac{n'^{\mu}}{2} + p_1'^+\frac{\bar{n}'^{\mu}}{2} + p_{1\perp}'^{\mu}, \nn \\
p_{1}^{\mu}&=p^-_1\frac{n^{\mu}}{2} + p^+_1\frac{\bar{n}^{\mu}}{2} + p_{1\perp}^{\mu}, \qquad
p_1^{\pm} = p_1^0 \mp p_1^3
\,,\end{align}
and similarly for $p_2$.
Here we chose $n^\mu=(1,0,0,1)$ and $\bar{n}^\mu=(1,0,0,-1)$, $z$ is the momentum fraction of the parton, $s$ the invariant mass of the jet, and $Q$ the center-of-mass energy of the $e^+e^-$ collision. The expression in the recoiled frame follows from the definition of $z$ and $s$ through $p_1'^- = z Q$ and $s=(p_1'+p_2')^2$, as well as $p_{1\perp}'^\mu = -p_{2\perp}'^\mu$ and the on-shell condition $p_1'^2 = p_2'^2=0$. Note that $|p_{i\perp}'|^{2}=z(1-z)s$.

The rotation between the two frames is described by 
\begin{equation}
\label{eq:rot}
\vec{p_{1}} = 
\begin{pmatrix} \cos\theta & 0 & -\sin\theta\\0 & 1 & 0\\\sin\theta & 0 & \cos\theta \end{pmatrix}
\vec{p_{1}}^{'}\,,
\end{equation}
implying $|p_{\perp}|^{2}=|p_{\perp}'|^{2}+\theta^{2}(p_1^3)^{2} - 2\theta \cos\phi' |p_{\perp}'| |p_1^3|$ in the small $\theta$ approximation, where $\phi'$ is the azimuthal angle around the $\vec n '$ axis. The large momentum components are the same in both frames, $p_i^- = p_i'^-$. The expression for the angularity $\tau_{n}$ becomes
\begin{align} \label{eq:4.3.3}
\tau_{n} &= \frac{1}{Q}\sum_{i}|p_{i\perp}|\biggl(\frac{p_i^+}{p_i^-}\biggr)^{\frac{b}{2}}
 = \frac{1}{Q} \sum_i \biggl(\frac{|p_{i\perp}|^{1+b}}{(p_i^-)^b}\biggr) \\
&= \frac{1}{(2Q)^{1+b}} \, z^{-b}\left(4z(1-z)s+(\theta Q)^{2} z^{2} - 4\theta \, Q \cos\phi' \, z^{\frac{3}{2}} \sqrt{(1-z)s} \,\right)^{\frac{1+b}{2}}\, \nn \\
 &\quad + \frac{1}{(2Q)^{1+b}} \, (1-z)^{-b}\, \left(4z(1-z)s+(\theta Q)^{2}(1-z)^{2} + 4\theta Q \, \cos\phi' (1-z)^{\frac{3}{2}} \sqrt{z s} \,\right)^{\frac{1+b}{2}}, 
\nn \end{align}
where $b>-1$.
Using the delta function trick (see \sec{dfs}), we switch to a cumulative measurement, writing the observable as 
\begin{align}
\label{eq:4.3.4}
M_{\text{obs}}&=\Theta[\tau_n^{\rm c}-\tau_{n}]
\,.\end{align}

Unfortunately is it not possible to invert \eq{4.3.4} to obtain an analytic solution for $s$ and subsequently extract the soft limit $z \to 0$. We can, however, use the power-law ansatz in \eq{sobs} to find the soft behavior of the observable. Since the equation is symmetric in $z \rightarrow  1-z$, we focus on finding the soft behavior in the $z \rightarrow 0$ limit.  Using 
\begin{align}
\label{eq:s0}
s|_{z\to0}= c_0(\phi)z^{-\alpha_0} \mu^2\,,
\end{align}
in \eq{4.3.3} and taking the $z\rightarrow 0$ soft limit, we find 
\begin{align} 
\label{eq:4.9}
\tau_n^c \Bigl(\frac{2Q}{\mu}\Bigr)^{1+b} &= z^{-b} \, \biggl(4\, c_0 \, z^{1-\alpha_0}+\Bigl(\frac{\theta Q}{\mu}\Bigr)^{2} \! z^{2} - 4\sqrt{c_0} \Bigl(\frac{\theta Q}{\mu}\Bigr) \cos\phi' \, z^{\frac{3-\alpha_0}{2}}  \,\Bigr)^{\frac{1+b}{2}}\, \nn \\
 &\quad + \biggl(4 \, c_0 \, z^{1-\alpha_0}+\Bigl(\frac{\theta Q}{\mu}\Bigr)^{2}+ 4\sqrt{c_0} \Bigl(\frac{\theta Q}{\mu}\Bigr)\, \cos\phi' (z)^{\frac{1-\alpha_0}{2}} \,\biggr)^{\frac{1+b}{2}}.
\end{align}
There is a single solution for $s$ in either of the soft limits and therefore this observable only has an upper boundary over the full range of $b$, i.e.~$c_0^- = 0$. The leading terms in \eq{4.9} are used to solve for $\alpha_0^+$ and $c_0^+$, and differ for $-1<b<0$, $b=0$ and $b>0$. We will analyze the last case in some detail and only provide the solutions for the others. 

Assuming $b>0$, the leading behavior in the $z\rightarrow 0$ limit  of \eq{4.9} is
\begin{align} 
\label{eq:leading terms}
 \tau_n^c \Bigl(\frac{2Q}{\mu}\Bigr)^{1+b} =  c_0^{\frac{1+b}{2}} \, z^{-b+(1-\alpha_0)(1+b)/2} + \left(\frac{\theta \, Q}{2\, \mu}\right)^{1+b},
\end{align}
and from this we infer  
\begin{align} 
\label{eq:sobs b>0}
c_0^+ =\frac{Q^2 (\tau_n^c)^{2/(1+b)} }{\mu^2} \bigl(1-k^{1+b}\bigr)^{\frac{2}{1+b}}\,, \qquad
\alpha_0^+ = \frac{1-b}{1+b}
\,,\end{align} 
where 
\begin{align}
\label{eq:def_k}
k\equiv \tfrac12 \theta \, (\tau_n^c)^{-1/(1+b)}
\,.\end{align}
Similarly, for $b=0$ we obtain
\begin{align} 
\label{eq:sobs b=0}
c_0^+ &=\frac{Q^2  (\tau_n^c)^{2/(1+b)}}{\mu^2}  \frac{1- k^2}{(2 + 2k\cos\phi)^2}\,, \qquad
\alpha_0^+ = 1
\,.\end{align}
For $-1<b<0$ the solution is a bit more difficult and reads 
\begin{align} 
\label{eq: sobs b<0}
  c_0^+ &=  \frac{Q^2 (\tau_n^c)^{2/(1+b)}}{\mu^2} \biggl[ 1 + k^2 \, \cos{2\phi}
   -  2 k |\cos \phi| \sqrt{1-k^2\sin^2{\phi}}  \,\biggr) \biggr]\,, \nn \\
   \alpha_0^+ &=1\,.
\end{align} 

In order to use \gojet, we rescale $s$ and choose an energy scale $\mu$. To be able to smoothly turn off the recoil, we choose $\mu$ in terms of the angularity, $\mu = Q\, (\tau_n^c)^{1/(1+b)}$. The only independent variable left is then given by $k$ in \eq{def_k}.
To be complete we also give the resulting observable input for \gojet: 
\begin{align} 
\label{eq:newobsrescaled}
\texttt{O} =& \, 1 - z^{-b}\left(z(1-z)s+
k^{2} z^{2} - 2k \cos\phi' \, z^{\frac{3}{2}} \sqrt{(1-z)s} \,\right)^{\frac{1+b}{2}}\, \nn \\
 &- (1-z)^{-b}\, \left(z(1-z)s+k^{2}(1-z)^{2} + 2k \cos\phi' (1-z)^{\frac{3}{2}} \sqrt{z s}\,\right)^{\frac{1+b}{2}}\,.
\end{align}

\begin{figure}[t]
\begin{subfigure}{0.47\textwidth}
\includegraphics[width=1\linewidth]{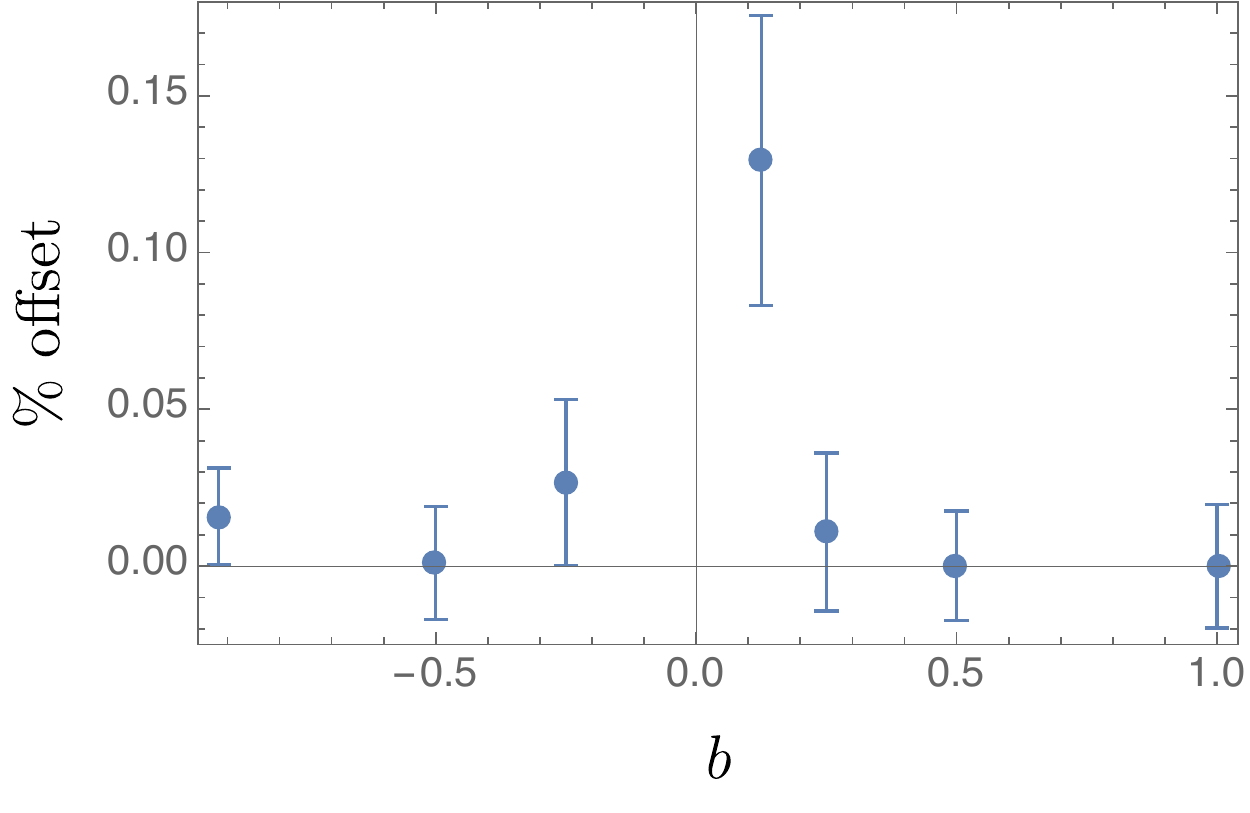}
\caption{}
\label{fig:errrec_a}
\end{subfigure}
\quad
\begin{subfigure}{0.48\textwidth}
\vspace{-2ex}
\includegraphics[width=1\linewidth]{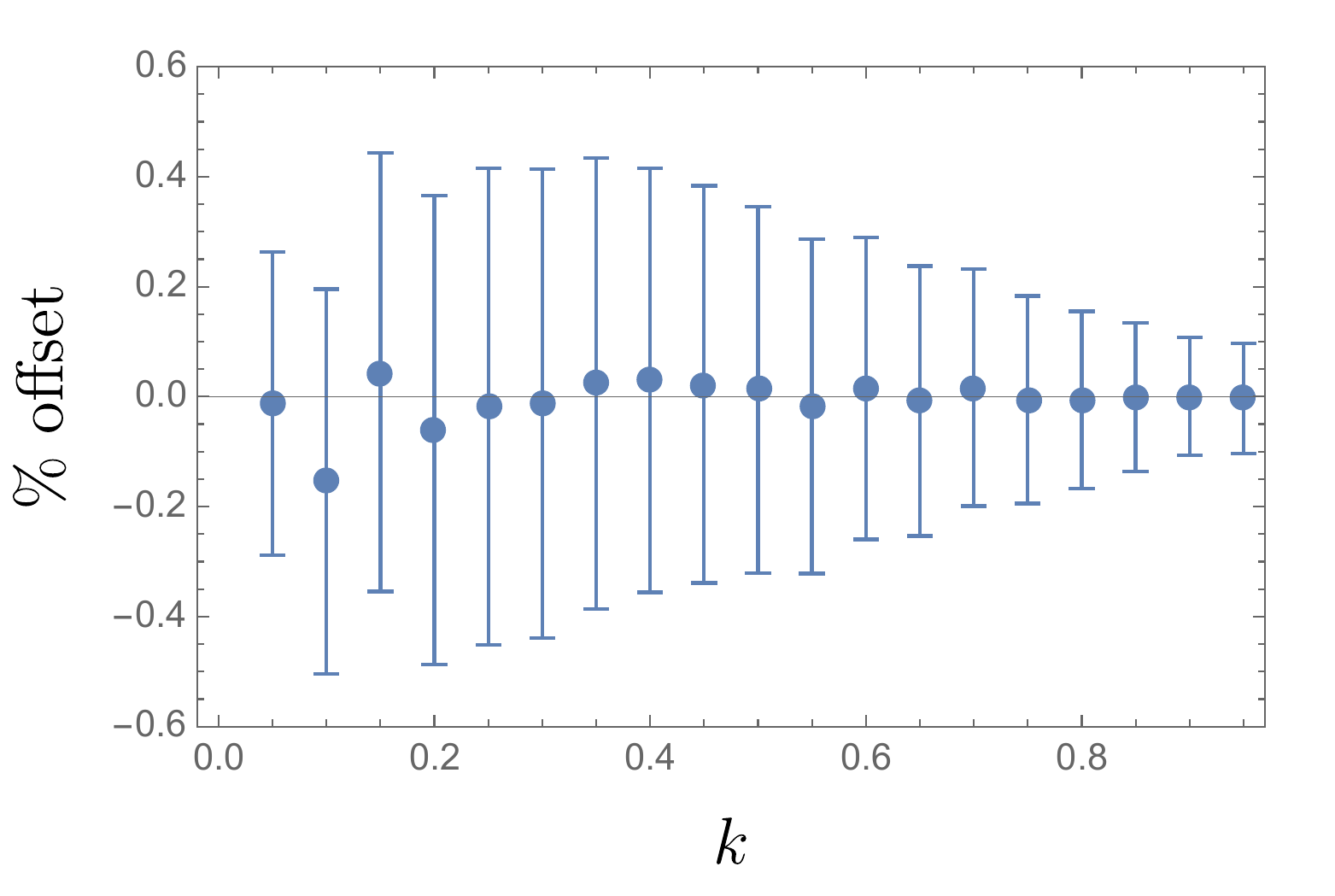}
\caption{}
\label{fig:errrec_b}
\end{subfigure}
\caption{The offset between our results for (a) different values of $b$ with $\theta=0$ and (b) different values of the recoil parameter $k$ with $b=0$ and the known results from the literature is shown.}
\label{fig:errrec} 
\end{figure}

\begin{figure}[t]
\begin{subfigure}{0.47\textwidth}
\includegraphics[width=1\linewidth]{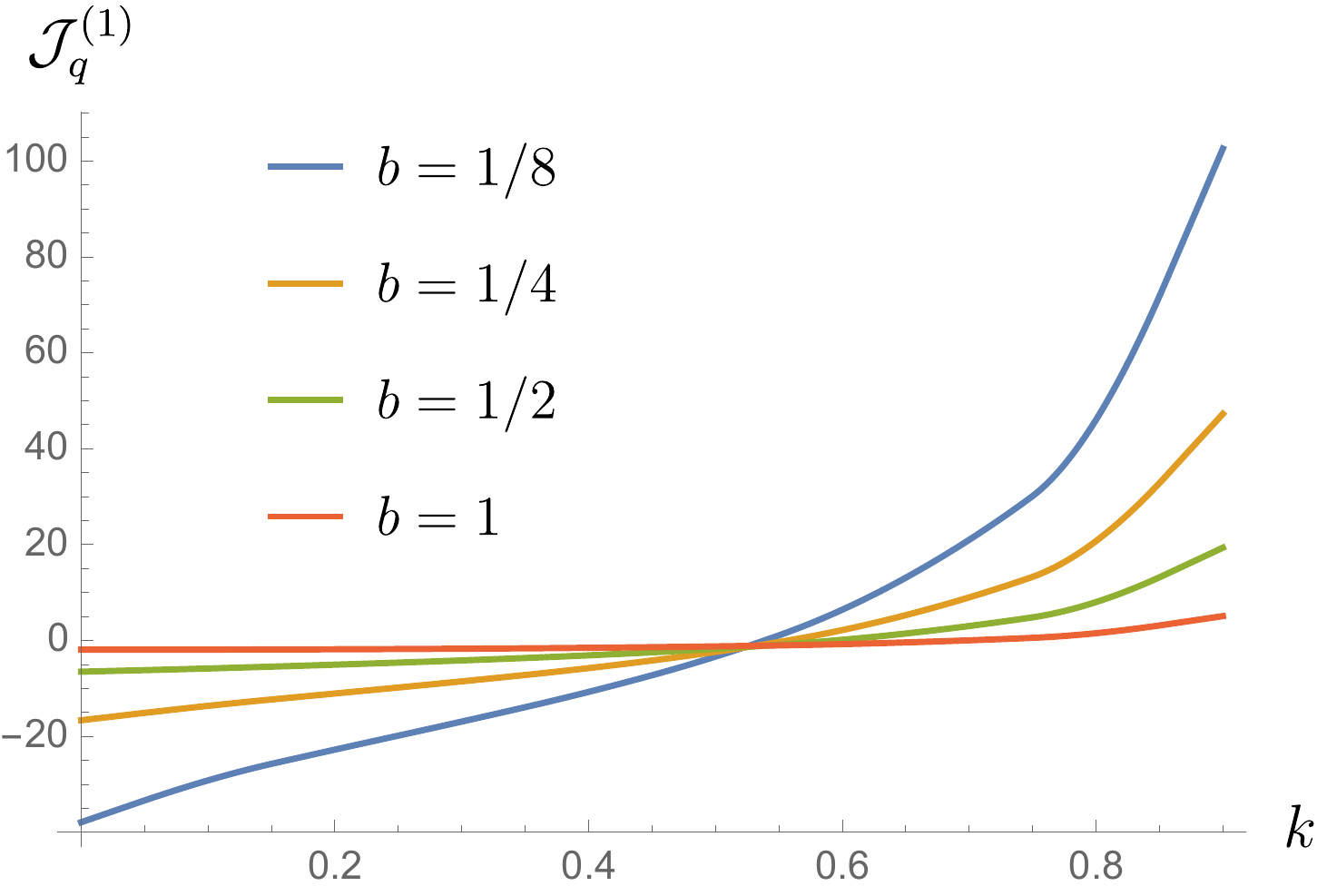}
\label{fig:resbigger} 
\end{subfigure}
\quad
\begin{subfigure}{0.47\textwidth}
\includegraphics[width=1\linewidth]{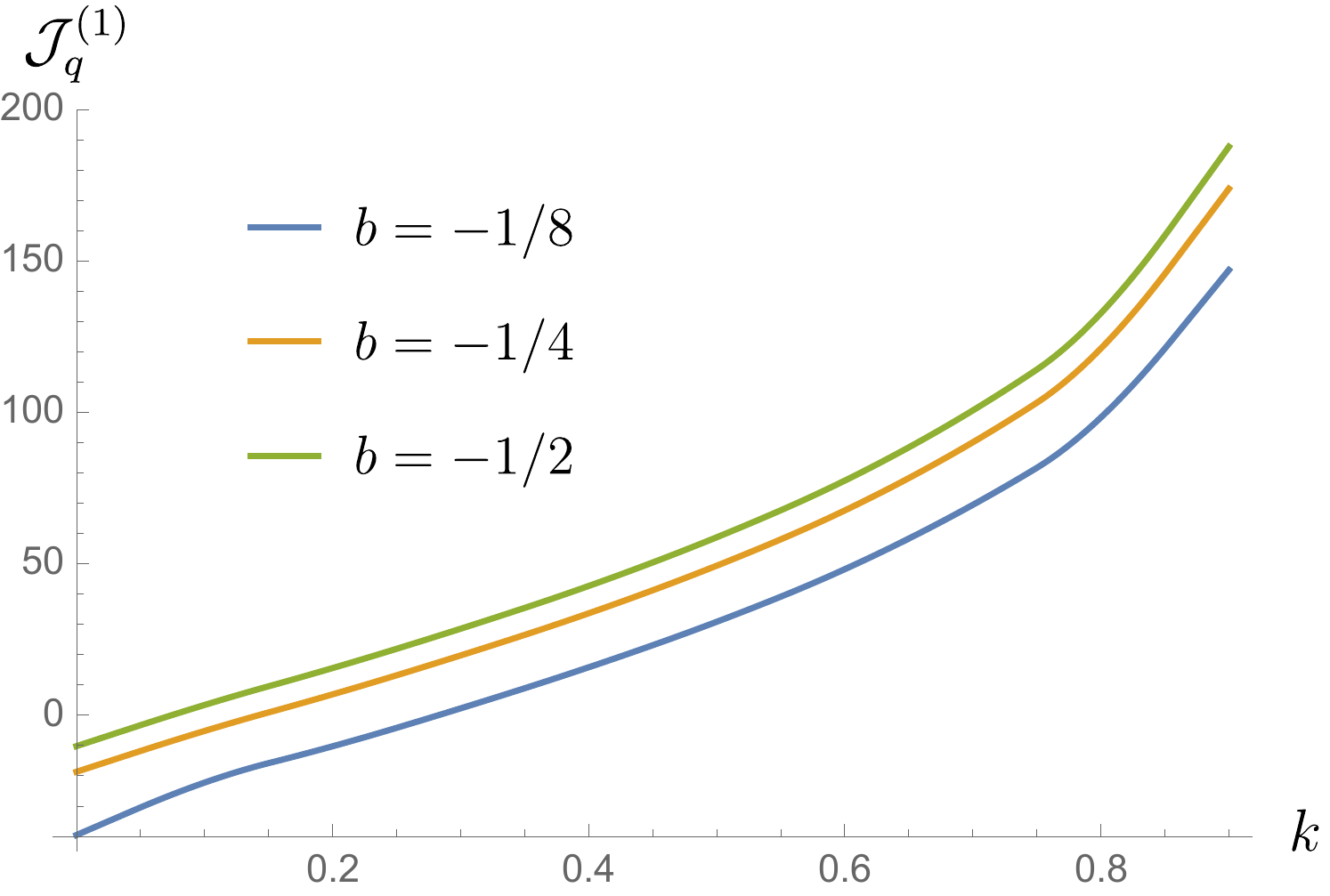}
\label{fig:ressmaller} 
\end{subfigure}
\caption{The results for the finite part of $\mathcal{J}_{q}^{1}$ for different values of $b$ as a function of $k$.}
\label{fig:resb}
\end{figure}

The jet function for $\theta = 0$ (without recoil) was calculated analytically in refs.~\cite{Hornig:2009vb,Budhraja:2019mcz} and we obtain the same results as can be seen in \fig{errrec_a}. 
The error bars indicate the uncertainty from our numerical integration. Ref.~\cite{Budhraja:2019mcz} includes a zero-bin subtraction~\cite{Manohar:2006nz} to avoid double counting with the soft function in their factorization, which we do not include. The zero-bin subtraction depends on the details of the factorization theorem (indeed it vanishes in ref.~\cite{Hornig:2009vb}), so we do not offer this as a standard functionality of \gojet.
The numerical integration for small values of $b$ is particularly challenging (as can be seen for $b=\tfrac{1}{8}$), because the sub-leading terms with respect to the leading soft behavior of the observable in \eq{sobs b>0} are particularly large in this case. A more detailed discussion of this issue and a method to cope with it is presented in \app{CTremap}. In \fig{errrec_b} we reproduce the known results for $b=0$ (broadening) and general recoil~\cite{Becher:2011dz}.  
Our new results for general $b$ including the effect of recoil, are shown in \fig{resb}. The error bars are not shown in this plot as they are negligibly small.

\subsection{Jet shape}
\label{sec:pedro}

\begin{figure}[t]
\begin{subfigure}{0.47\textwidth}
\includegraphics[width=1\linewidth]{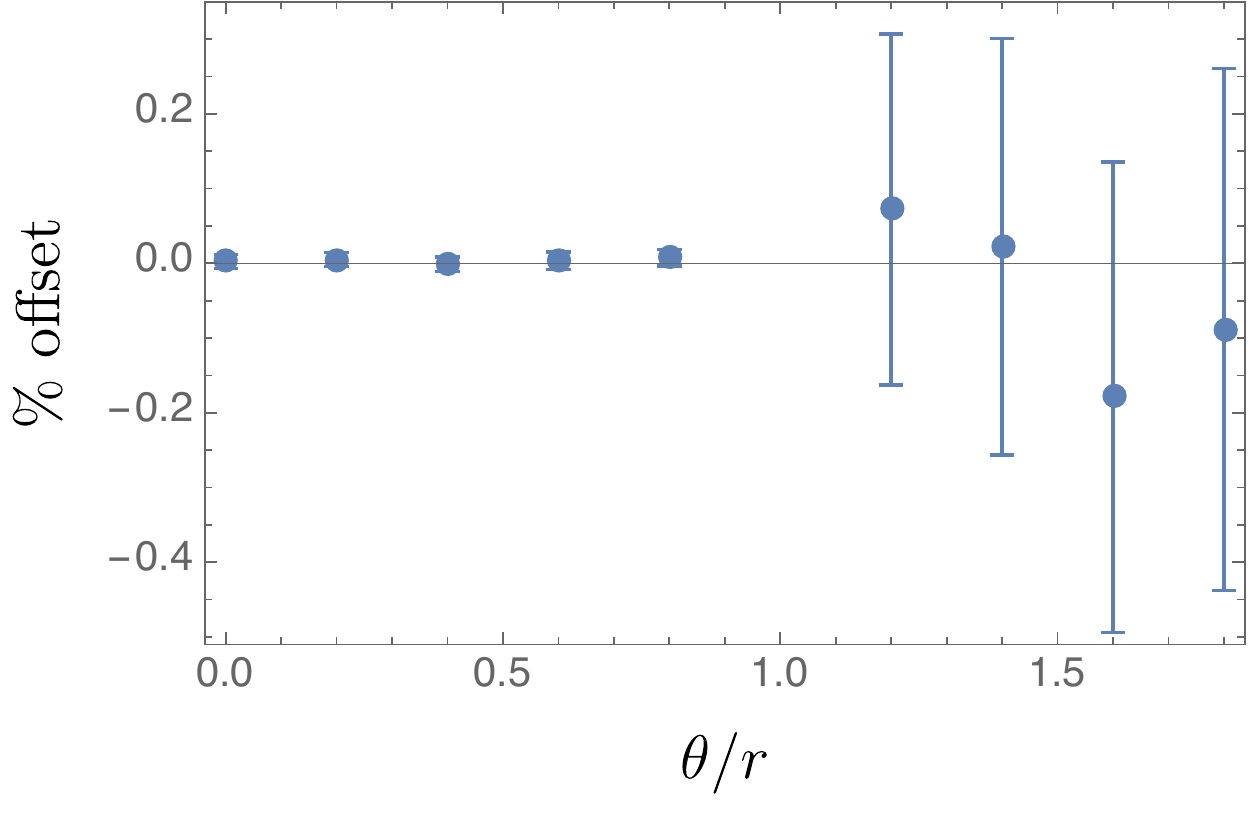}
\caption{}
\label{fig:pedro_a}
\end{subfigure}
\quad
\begin{subfigure}{0.47\textwidth}
\vspace{-0.4ex}
\includegraphics[width=1\linewidth]{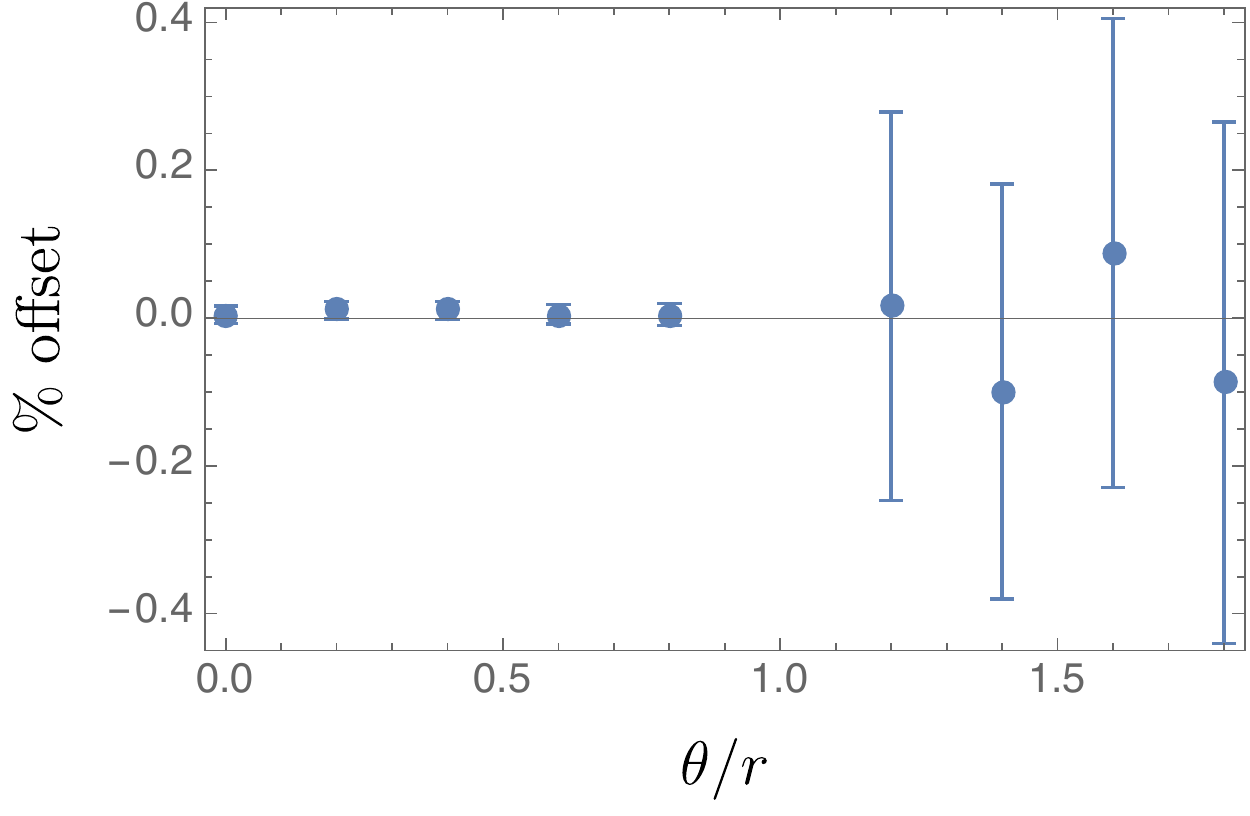}
\caption{}
\label{fig:pedro_b}
\end{subfigure}
\caption{The offset between our finite result of the (a) gluon and (b) quark jet function and \cite{Cal:2019hjc} for several values of $\frac{\theta}{r}$.}
\label{fig:pedroerror} 
\end{figure}

As another nontrivial example, we calculate the jet function for the classic jet shape observable, reproducing the one-loop result of ref.~\cite{Cal:2019hjc}. The jet shape describes the average energy fraction $z_r$ inside a cone of angular size $r$ around the jet axis. As in \sec{new}, recoil from soft radiation displaces the jet axis from the initial parton by an angle $\theta$. This breaks the azimuthal symmetry, requiring one to integrate over $\phi$. We have checked that our poles match exactly with the poles in \cite{Cal:2019hjc} for all values of $\theta$ and $r$. The difference between the finite term is always below 0.5\%. This has been illustrated in \fig{pedro_a} for gluon jets and \fig{pedro_b} for quark jets. We note that run time is not an issue, as less precision is needed in phenomenological results and the distribution can be interpolated. 
Our calculation represents the second independent calculation of this observable and thereby delivers a useful cross check of the results of ref.~\cite{Cal:2019hjc}.

\section{Conclusions}
\label{sec:conc}

In this paper we developed an automated approach for calculating one-loop jet functions, and provide an implementation in the accompanying \Mathematica package called \gojet. We use geometric subtraction~\cite{Herzog:2018ily} to isolate the soft and collinear singularities. The collinear counterterm does not depend on the details of the observable, except that certain observables do not require it. We find that the soft counterterm depends on the behavior of the observable in the soft limits, which can be described by a power law. While the user must provide \gojet with this power law as input, we present a strategy to extract this in a highly nontrivial example. 
We employed cumulative distributions, such that observables correspond to integrating over certain regions of phase space, and thereby avoiding plus distributions. 
We have demonstrated our approach by reproducing the known one-loop jet function for a range of observables, and calculating, for the first time, the jet function for angularities including recoil. For broadening ($b=0$ in our conventions) the effect of recoil must be kept~\cite{Becher:2012qc}, while for $b>0$ it is formally power suppressed but can be numerically large~\cite{Budhraja:2019mcz}. For $b$ close to 0, we encountered numerical convergence issues, due to an integrable divergence. We addressed this problem by substantially improving the counterterm through a remapping.

Our approach focusses on IR-safe observables, and we did not address the IR-unsafe case. Jet functions containing IR divergences are sensitive to nonperturbative physics, and our purely partonic calculation must be supplemented by a (universal) nonperturbative function that subtracts these divergences. A prime example is initial-state jets, which are described by beam functions~\cite{Stewart:2009yx}. Beam functions contain infrared divergences, which are removed by matching onto parton distribution functions, leaving finite matching coefficients.

The automated approach and code presented here provides a very useful tool, calculating jet functions at one-loop order. Very few two-loop jet functions are known, and an automated approach would allow many resummation calculations to be extended to NNLL$'$ or N$^3$LL accuracy. At this order the singular limits become more complicated, the order of subtractions matter, and the parametrization of the observable in these limits will no longer be a simple power law, complicating the counterterms.

\acknowledgments
This work is supported by the ERC grant ERC-STG-2015-677323, the NWO projectruimte 680-91-122, the NWO Vidi grant 680-47-551, the UKRI FLF grant Mr/S03479x/1 and the D-ITP consortium, a program of NWO that is funded by the Dutch Ministry of Education, Culture and Science (OCW). 


\appendix

\section{$G_{2}$ Subtraction Term for Rapidity Divergences}
\label{app:rap}
When the soft limit of the observable scales as $1/z$, we need a rapidity regulator to control the singularities. The resulting expressions for $G_{2}$ with rapidity regulator are given  by 
 \begin{align}
\label{eq:g2rap1}
G_{q,2} &=\frac{2C_{F}}{\epsilon}  \frac{e^{\gamma_E\epsilon}  }{ \sqrt{\pi}\,\Gamma(\tfrac{1}{2}-\epsilon)}\bigg(\frac{\nu}{\omega}\bigg)^{\eta}\int_0^\pi\! \df \phi\, \Theta(\Phi)(\sin\phi)^{-2\eps}  \nn\\
&\qquad\bigg[\frac{(c_1^+)^{-\epsilon}}{\eta +\eps(1-\alpha_1^+)} A^{-\eta-\eps(1-\alpha_1^+)}- \frac{(c_1^-)^{-\epsilon}}{\eta +\eps(1-\alpha_1^-)} A^{-\eta-\eps(1-\alpha_1^-)} \bigg] \, ,
     \nn \\
G_{g,2} &=\, \frac{C_{A}}{ \epsilon}  \frac{e^{\gamma_E\epsilon}  }{ \sqrt{\pi}\,\Gamma(\tfrac{1}{2}-\epsilon)}\bigg(\frac{\nu}{\omega}\bigg)^{\eta}\int_0^\pi\! \df \phi\, \Theta(\Phi)(\sin\phi)^{-2\eps}   \nonumber \\
&\qquad  \bigg[\frac{(c_0^+)^{-\epsilon}}{\eta +\eps(1-\alpha_0^+)} A^{-\eta-\eps(1-\alpha_0^+)}- \frac{(c_0^-)^{-\epsilon}}{\eta +\eps(1-\alpha_0^-)} A^{-\eta-\eps(1-\alpha_0^-)}   \nn\\
&\qquad + \frac{(c_1^+)^{-\epsilon}}{\eta +\eps(1-\alpha_1^+)} A^{-\eta-\eps(1-\alpha_1^+)}- \frac{(c_1^-)^{-\epsilon}}{\eta +\eps(1-\alpha_1^-)} A^{-\eta-\eps(1-\alpha_1^-)} \bigg].
 \end{align}

\section{Counterterm Mapping}
\label{app:CTremap}

In this appendix we discuss how to improve the convergence of the soft subtraction through a mapping. For simplicity, we consider only the soft singularity at $z=0$, for which the finite term generated by the geometric subtraction is of the form:
\begin{equation}
\label{eq:APPfiniteterm}
\int_0^1\! \df z\,\Big[\frac{f(z)\Theta(O(z))-f(0)\Theta(O_0(z))}{z}\Big]
\,.\end{equation}
Here we suppressed the dependence (and integrals) over $s$ and $\phi$, extracting the $1/z$ singularity from the integrand $Q$, i.e.~$f = z Q$.  While this integrand is by construction integrable, poor numerical convergence may be caused by mismatch of the observable $O$ and its soft limit $O_0$. This problem can become particularly severe if $O(z)$ has a fractional power series in $z$, as we illustrate below. 

To improve the convergence of the integral, we apply the following mapping (to the counterterm only):
\begin{equation}
G\!: z\to\frac{z+g(z)}{1+g(z)} \,.
\end{equation}
This maps the interval $0\le z\le1$ onto itself, as long as $z+g(z)>0$, and the subtracted integral will remain the same as long as the function $g(z)$ decreases faster near $z=0$ than $z$ itself, i.e., it satisfies
\begin{equation}
\lim_{z\to 0}\frac{g(z)}{z}=0\,.
\end{equation}
Applying this map, we can replace \eq{APPfiniteterm} with:
\begin{equation}
\label{eq:APPRfiniteterm}
\int_0^1\! \df z\,\Big[
\frac{f(z)\Theta(O(z))}{z}
-\frac{f(0)\Theta(O_0(G(z)))}{G(z)}\Big|\frac{\partial G(z)}{\partial z}\Big|\Big].
\end{equation}
One can now construct the function $g(z)$ to map $O_0(G(z))$ closer to $O(z)$ in the region $z\to0$. 

\begin{figure}[t]
\centering
\includegraphics[width=0.5\textwidth]{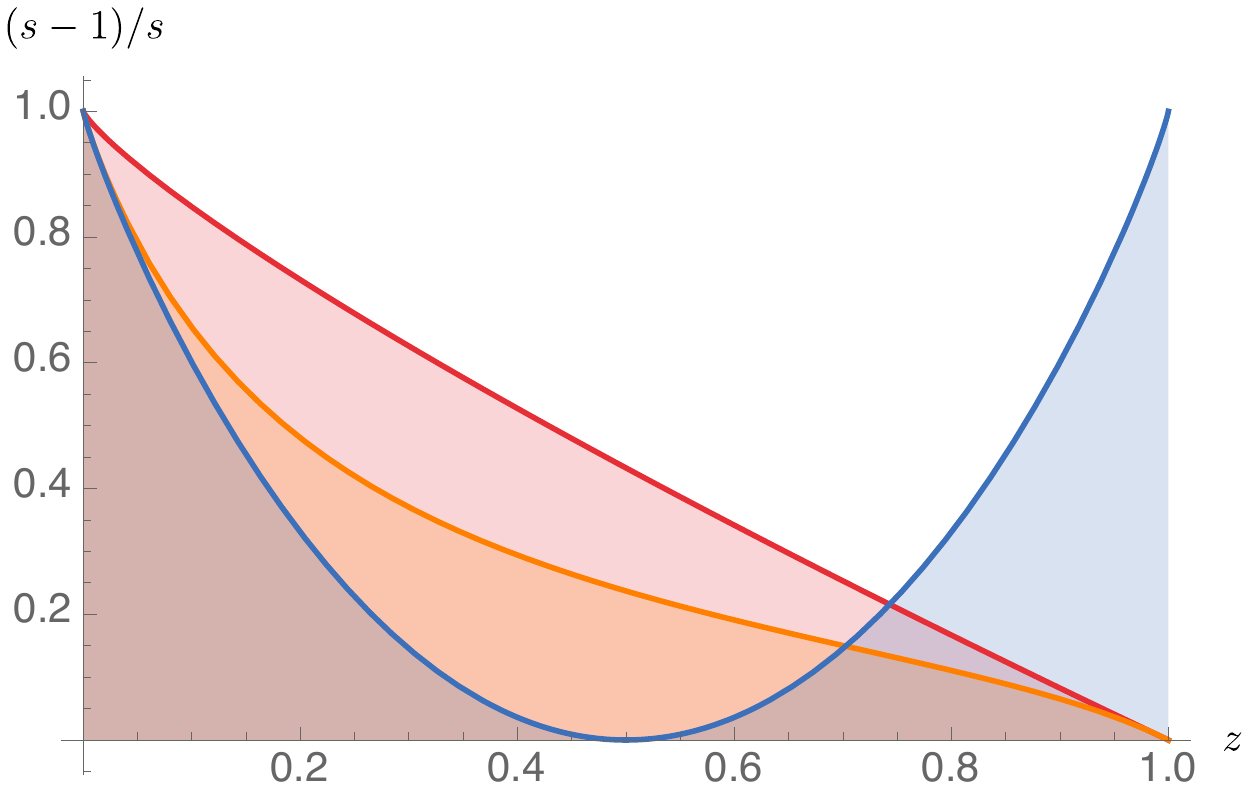}
\caption{The plot shows the observable (blue), its soft approximation in \eq{exO0} (red) and the remapped soft approximation in \eq{exOremap} (orange).}
\label{fig:CTmap} 
\end{figure}

For the angularities with recoil in \sec{new}, we encounter the following instructive example
\begin{equation}
O(z)=1/s-\frac{(z(1-z))^{\frac{b-1}{b+1}}}{(z^b+(1-z)^b)^{\frac{2}{1+b}}}\,,
\end{equation}
which has poor convergence for small positive values of $b$. Already $b=1/10$ yields a sufficiently challenging scenario, for which 
the power series around $z=0$ is given by:
\begin{align}
\label{eq:exOmany}
O(z)&=1/s-
{z}^{{\frac {9}{11}}}+{\frac {20}{11}{z}^{{\frac {101}{110}}}}+{\frac 
{90}{121}{z}^{{\frac {56}{55}}}}-{\frac {60}{1331}{z}^{{\frac {123}{
110}}}}+{\frac {195}{14641}{z}^{{\frac {67}{55}}}}-{\frac {936}{161051
}{z}^{{\frac {29}{22}}}}\nn\\
& \quad
+{\frac {5460}{1771561}{z}^{{\frac {78}{55}}}}
-{\frac {35880}{19487171}{z}^{{\frac {167}{110}}}}+{\frac {255645}{
214358881}{z}^{{\frac {89}{55}}}}-{\frac {1931540}{2357947691}{z}^{{
\frac {189}{110}}}}\nn\\
& \quad
-{\frac {25922165435}{25937424601}{z}^{{\frac {20}{
11}}}}-{\frac {5136983395938}{3138428376721}{z}^{{\frac {211}{110}}}}
+\mathcal{O}(z^2)\,.
\end{align}
It is thus apparent that the leading term approximation
\begin{align}
\label{eq:exO0}
O_0(s,z,\phi)&=1/s-{z}^{{\frac {9}{11}}}
\end{align}
gives only a poor approximation of the full result. Substituting $z=G(z)$ with 
\begin{equation}
\label{eq:exOremap}
g(z)=z\sum_{i=1}^{11} c_i \,z^{\frac{i}{10}}
\end{equation}
into \eq{exO0} we can match \eq{exOmany} by iteratively solving for the constants $c_i$.
This procedure yields:
\begin{align}
&{ c_1}={\frac {20}{9}},\quad { c_2}={\frac {110}{81}},\quad { c_3}={\frac {220}{2187}},\quad { c_4}=-{\frac {385}{19683}}\,,\quad 
{ c_5}={\frac {1232}{177147}},\quad { c_6}=-{\frac {15400}{4782969}},\nn\\
&{ c_7}={\frac {74800}{43046721}}\,,\quad
{ c_8}=-{\frac {402050}{387420489}},\quad
{ c_9}={\frac {20906600}{31381059609}},\quad
 c_{10}=-{\frac {345319185959}{282429536481}},\nn\\
& c_{11}=-{\frac {6338162484818}{2541865828329}}\,.
\end{align}
The resulting curves are plotted in \fig{CTmap}, highlighting the improvement due to the remapping. A \Vegas run using $5\cdot10^9$ points for the finite part of the quark jet function of this observable yields $-48.63(2)$ without the mapping, while we obtain $-48.745(9)$ after the mapping. The true value is $-48.7731$, indicating that the remapped counterterm yields a result significantly closer to the true value. In both cases it becomes clear that the offset is not completely covered by the uncertainty. While the remapping may thus improve convergence, it may not completely solve the issue.

\section{Azimuthal Integral}
\label{app:azimuth}

In this appendix we evaluate the integral
\begin{equation}
I(a,b;\eps)=\int_a^b\! \df \phi\, (\sin\phi)^{-2\eps} 
\,.\end{equation}
One can convert this integral into a Gauss-type hypergeometric integral using the transformation $\cos\phi=1-2x$.
However this leads to square roots in the denominator which do not naively lead to a polylogarithmic expression.
Instead, one can rewrite the integral as a contour integral in the complex plane
using the transformation $z=e^{i\phi}$, such that
\begin{equation}
\sin\phi=\frac{z^2-1}{2\img z} 
\,,\end{equation}
leading to the following representation
\begin{equation}
I(a,b;\eps)=-\img \int_{e^{\img a}}^{e^{\img b}} \frac{\df z}{z} \Big(\frac{z^2-1}{2\img z} \Big)^{-2\eps} 
\,.\end{equation}
The integrand can be chosen to have branch cuts on the real axis for $z<0$ and for $z>1$. For $0<a,b<\pi$, which is the range of physical interest, no branch cuts are ever crossed. 

It is convenient to perform the integral on a contour along the real axis from $0<z<A$ with $0<A<1$, i.e.,
\begin{equation}
F(A;\eps)=-\img 2^{2\eps} e^{-\img \pi\eps}\int_{0}^{A} \frac{\df z}{z} \Big(\frac{1-z^2}{z} \Big)^{-2\eps}. 
\end{equation}
The result can analytic continued to the case of interest with $A=e^{\img a}$. 
We then obtain (in essence via the residue theorem)
\begin{equation} \label{eq:ItoF}
 I(a,b;\eps)=F(e^{ia};\eps)-F(e^{ib};\eps)\,.
\end{equation}
While the divergence at $z=0$ requires careful treatment, this drops out in the difference of the two terms in \eq{ItoF}. 
We performed the integral using the Maple package Hyperint \cite{Panzer:2014caa}, finding that the integral can performed order by order in $\epsilon$ in terms of harmonic polylogarithms. This is to be expected, given that its singularities are located at $z=0,-1,1$. Up to order $\eps^2$ we can express the result in terms of the classical polylogarithms:
\begin{equation}
I(a,b;\eps)=\sum_{n=0}^\infty I^{(n)}(a,b)\eps^n
\end{equation}
with 
\begin{align}
I^{(0)}(a,b)&=b-a\,, \nn\\
I^{(1)}(a,b)&= 2\img\, \Li_2(e^{\img a})-2\img\, \Li_2(e^{\img b})+2\img\, \Li_2(-e^{\img a})-2\img\, \Li_2(-e^{\img b})+\img (a-b)(-a+\pi-b)\nn\\
&\quad+(-2a+2b)\ln2\,, \nn \\
I^{(2)}(a,b)&=-\tfrac{2}{3}\img \ln^3(e^{\img b}+1)-2\img \ln^2(e^{\img b}+1)\ln(1-e^{\img b})-4b\ln(e^{\img b}+1)\ln(1-e^{\img b})\nn\\
&\quad+2\img \ln^2(e^{\img a}+1)\ln(1-e^{\img a})+4a\ln(e^{\img a}+1)\ln(1-e^{\img a})-2\img \ln(e^{\img b}+1)\ln^22\nn\\
&\quad-2\img \ln(1-e^{\img b})\ln^22+2\img \ln(e^{\img a}+1)\ln^22+2\img \ln(1-e^{\img a})\ln^22-4\img\, \Li_3(e^{\img a})\nn\\
&\quad+4\img\, \Li_3[-(-1+e^{\img a})/(e^{\img a}+1)]-4\img\, \Li_3[-(-1+e^{\img b})/(e^{\img b}+1)]+2a \ln^2(e^{\img a}+1)\nn\\
&\quad-2b \ln^2(1-e^{\img b})-2b \ln^2(e^{\img b}+1)+4\img\, \Li_3(\tfrac12 +\tfrac12 e^{\img a})-8\img\, \Li_3[1/(e^{\img a}+1)]\nn\\
&\quad+4\img\, \Li_3(\tfrac12 -\tfrac12 e^{\img a})-8\img\, \Li_3(1-e^{\img a})+4\img\, \Li_3(e^{\img b})-4\img\, \Li_3(\tfrac12 -\tfrac12 e^{\img b})\nn\\
&\quad+8\img\, \Li_3(1-e^{\img b})+4\img\, \Li_3(-e^{\img b})+8\img\, \Li_3(1/(e^{\img b}+1))-4\img\, \Li_3(\tfrac12 +\tfrac12 e^{\img b})\nn\\
&\quad+2(b-a)\ln^22+2\img (a-b)(-a+\pi-b)\ln2-4\img \ln(e^{\img a}+1)\ln(1-e^{\img a})\ln2\nn\\
&\quad+(2\pi-4a)\Li_2(-e^{\img a})+\tfrac{1}{3}\img \pi^2\ln(1-e^{\img a})+4\img \ln2\,\Li_2(-e^{\img a})\nn\\
&\quad-4\img \ln2\,\Li_2(e^{\img b})+4\img \ln(e^{\img b}+1)\Li_2(e^{\img b})+4\img \ln(1-e^{\img b})\Li_2(e^{\img b})\nn\\
&\quad-4\img \ln(e^{\img a}+1)\Li_2(e^{\img a})-4\img \ln(1-e^{\img a})\Li_2(e^{\img a})+4\img \ln2\,\Li_2(e^{\img a})\nn \\
&\quad+\tfrac{1}{6}(a-b)(3\pi^2-6\pi a-6\pi b+4a^2+4ab+4b^2)+\tfrac{2}{3}\img \ln^3(e^{\img a}+1)\nn\\
&\quad+(-2\pi+4b)\Li_2(e^{\img b})+(2\pi-4a)\Li_2(e^{\img a})-\tfrac{1}{3}\img \pi^2\ln(1-e^{\img b})\nn\\
&\quad+4\img \ln(e^{\img b}+1)\ln(1-e^{\img b})\ln2+4\img \ln(e^{\img b}+1)\Li_2(-e^{\img b})\nn\\
&\quad+4\img \ln(1-e^{\img b})\Li_2(-e^{\img b})-4\img \ln2\,\Li_2(-e^{\img b})+\img \pi^2\ln(e^{\img b}+1)\nn\\
&\quad-\img \pi^2\ln(e^{\img a}+1)-4\img \ln(e^{\img a}+1)\Li_2(-e^{\img a})-4\img \ln(1-e^{\img a})\Li_2(-e^{\img a})\nn\\
&\quad+2a\ln^2(1-e^{\img a})-4\img\, \Li_3(-e^{\img a})+(-2\pi+4b)\Li_2(-e^{\img b})
\,.\end{align}

\newpage
\bibliographystyle{JHEP}
\bibliography{biblio}

\end{document}